\documentclass[aip,rsi,amsmath,amssymb,reprint,floatfix,nofootinbib]{revtex4-2}

\usepackage{graphicx}
\usepackage{dcolumn}
\usepackage{bm}
\usepackage{xfrac}
\usepackage[T1]{fontenc}
\usepackage[utf8]{inputenc}
\usepackage[dvipsnames]{xcolor}

\usepackage{tikz}
\usetikzlibrary{arrows,decorations.markings}
\usetikzlibrary{arrows.meta}
\usepackage{tikz-3dplot}
\usetikzlibrary{math} 
\usetikzlibrary{decorations.pathmorphing, patterns,shapes}
\usetikzlibrary{calc}
\usepgflibrary{arrows}
\tikzset{
	partial ellipse/.style args={#1:#2:#3}{
		insert path={+ (#1:#3) arc (#1:#2:#3)}
	}
}
\usepackage{circuitikz}

\usepackage{siunitx}
\usepackage{hyperref}
\hypersetup{
    colorlinks=true, 
    linktoc=all,     
    linkcolor=blue,  
    citecolor=magenta,
    urlcolor=blue	
}
\usepackage{url}
\usepackage[nooneline,raggedright]{subfigure}
\usepackage{tikz}
\newcommand*\circled[1]{\tikz[baseline=(char.base)]{
    \node[shape=circle, draw, inner sep=1pt, 
        minimum height=12pt] (char) {#1};}}

\newcommand{\dd}{\text{d}}
    
\definecolor{amber}{rgb}{1.0, 0.75, 0.0}
\definecolor{alizarin}{rgb}{0.82, 0.1, 0.26}
\definecolor{ballblue}{rgb}{0.13, 0.67, 0.8}

\begin{document}

\title{Compact beam position monitor using a segmented toroidal coil}

\author{F.\,Abusaif}  
\affiliation{Physics Institute III B, RWTH Aachen University, 52074, Aachen, Germany}
\affiliation{Institute for Nuclear Physics, Forschungszentrum J\"ulich, 52425, J\"ulich, Germany}
\affiliation{present address: Institute for Beam Physics and Technology, Karlsruhe Institute of Technology, 76131, Karlsruhe, Germany}

\author{F.\,Hinder}
\affiliation{Physics Institute III B, RWTH Aachen University, 52074, Aachen, Germany}
\affiliation{Institute for Nuclear Physics, Forschungszentrum J\"ulich, 52425, J\"ulich, Germany}

\author{A.\,Nass}
\affiliation{Institute for Nuclear Physics, Forschungszentrum J\"ulich, 52425, J\"ulich, Germany}

\author{J.\,Pretz}
\affiliation{Physics Institute III B, RWTH Aachen University, 52074, Aachen, Germany}
\affiliation{Institute for Nuclear Physics, Forschungszentrum J\"ulich, 52425, J\"ulich, Germany}

\author{F.\,Rathmann}
\affiliation{Institute for Nuclear Physics, Forschungszentrum J\"ulich, 52425, J\"ulich, Germany}
\affiliation{present address Brookhaven National Laboratory, Upton, New York, 11973, USA}
\author{H.\,Soltner}
\affiliation{Central Institute of Engineering, Electronics and Analytics, Forschungszentrum Jülich, 52425 Jülich, Germany}

\author{D.\,Shergelashvili}
\affiliation{High Energy Physics Institute, Tbilisi State University, 0186 Tbilisi, Georgia}

\author{R.\,Suvarna}
\affiliation{Physics Institute III B, RWTH Aachen University, 52074, Aachen, Germany}
\affiliation{GSI Helmholtz Centre for Heavy Ion Research, 64291 Darmstadt, Germany}

\author{F.\,Trinkel}
\affiliation{Physics Institute III B, RWTH Aachen University, 52074, Aachen, Germany}
\affiliation{Institute for Nuclear Physics, Forschungszentrum J\"ulich, 52425, J\"ulich, Germany}
\affiliation{present address: German Aerospace Center (DLR), Linder H\"ohe, 51147 K\"oln, Germany}

\date{\textbf{for the Editorial team}, \today}

\begin{abstract}
A new, compact beam position monitor based on segmented a toroidal coil surrounding the charged particle beam has been investigated. It makes use of the induced voltages in the windings instead of the induced charge imbalance on capacitor plates in the conventional beam position monitors. We theoretically investigate the response of the coils to the bunched particle beam based on a lumped element model and compare it with measurements in the laboratory and in the storage ring COSY, in terms of beam displacement. As to the frequency response of the coils, we find a resonant behavior, which may be exploited to increase the sensitivity of the device further. The resolution presently achieved is about
\SI{5}{\micro m} in a \SI{1}{s} time interval for a beam current of
\SI{0.5}{mA}. 
\end{abstract}

\maketitle
\tableofcontents

\section{Introduction}
\label{sec:introduction}

Experiments searching for Electric Dipole Moments (EDMs) of charged particles using storage rings are at the forefront of the incessant quest to find new physics beyond the Standard Model (SM). These investigations bear the potential to shed light on the origin of the hitherto unexplained large matter-antimatter asymmetry in the Universe\,\cite{POSPELOV2005119,robson:2018}. The combined predictions of the SM of particle physics and of cosmology fall short of the experimentally observed asymmetry by about seven to eight orders of magnitude\,\cite{Bernreuther2002}. 

The JEDI collaboration (Jülich Electric Dipole moment Investigations, see \url{http://collaborations.fz-juelich.de/ikp/jedi})  is currently leading the effort to scrutinize the technical feasibility of the storage ring approach to the determination of the EDMs of protons\,\cite{cpEDM2021}. In the framework of systematic beam and spin dynamics studies, a dedicated experiment to determine the deuteron EDM\cite{Rathmann:2013rqa,Morse:2013hoa,PhysRevAccelBeams.23.024601}  is presently being carried out at the storage ring COSY in Jülich (Cooler Synchrotron COSY at Forschungszentrum Jülich, Germany\,\cite{Maier19971,PAX}).

Given the extremely small anticipated values of the EDMs of the charged particles, down to \SI{e-29}{e.cm}, the control of systematic effects in the ring is of paramount importance, and high-precision monitoring of the positions of the beams in the ring constitutes one of the great challenges in these experiments.  This entails precise control of the beam positions along the orbit in the machine. In order to improve the knowledge about the absolute beam orbit in COSY, various alignment campaigns were conducted, during which all magnetic elements in the machine were positioned to about \SI{0.2}{\milli \meter}, or \SI{0.2}{\milli \radian}, respectively. In addition, absolute beam-offset parameters for each of the installed BPMs were obtained from a dedicated beam-based alignment effort, as described in~\cite{Wagner:2020akw}.

The beam position monitors (BPMs) used at COSY are capacitive measuring
devices~\cite{wendt2020bpm,Maier19971}. The position of the beam is determined based on the induced charge
imbalance on opposing capacitor plates. The typical insertion length of such a
pair of BPMs along the beam direction, providing horizontal $x$ and vertical
$y$ positions, amounts to about \SI{500}{mm}. For beams comprising
approximately \SI{e9}{particles}, capacitive BPMs provide a resolution of a few 
\SI{}{\micro\meter}  for a measurement time of \SI{1}{\second} and an
accuracy of about \SI{100}{\micro \meter}. 

These \textit{capacitive} BPMs, however, appear unattractive because their length limits the number of devices that can actually be installed in a machine. It is therefore imperative to develop alternative BPMs that require significantly less installation space. 
In our case, it was not possible to install conventional BPMs
in front of and behind a one-meter-long radio frequency  Wien filter~\cite{Slim2016116}, essential for the EDM experiments.
Therefore, a newly developed device was used 
to determine the beam position during the EDM measurements.

In this publication, we describe an approach based on the \textit{induction} of voltages in opposing segments of a compact toroidal coil through which the beam passes, providing a position value based on the measured voltage imbalance. 
Previous studies of our group are discussed in proceedings~\cite{Hinder:2016hqi} and in a PhD thesis~\cite{phd_abusaif}.
Prior to our work, these so-called Rogowski coils 
for determining the beam current and the beam position were discussed in reference~\cite{berners_rogowski}.
In this document, we will discuss the theoretical description and the resonant behavior in detail.

One advantage of induction coils for this particular application,
apart from their mechanical simplicity,
is that they offer a large sensor surface in small volume due to the
large number of windings.
In addition they can be operated in resonance to increase the induced voltage and
that -- due to the induction principle --
they benefit from the high revolution frequency of the beam.
Furthermore it should be noted that BPMs based on induction coils are
sensitive to the time derivative of the beam current $\dot{I}$, while
capacitive BPMs are sensitive to $I$. Unlike BPM systems based on SQUIDs~\cite{Haciomeroglu:2018son}, they do not necessarily
require low temperatures or vacuum conditions during testing and beam
operation, which facilitates their development in the laboratory. 
Because of their short insertion length of about \SI{5}{cm}, Rogowski-type
BPMs appear ideally suited for a future dedicated EDM storage ring, see
Chaps.\,7 and 8 of Ref. \cite{cpEDM2021}. 

This paper is organized as follows. In Sec.\,\ref{sec:coils-as-bpms}, the
basic principles of BPMs based on toroidal coils are discussed.
In Sec.\,\ref{sec:experimental-setup}, the technical realization of the BPM,
the experimental test stand, as well as the calibration procedure are described.
Sec.\,\ref{sec:investigations-test-stand} summarizes the results of the
investigations carried out at the test stand. The results of the measurements
obtained after installation in the COSY storage ring are described in
Sec.\,\ref{sec:installation-COSY}, followed by the conclusions and outlook in Sec.~\ref{sec:conclusion}. 
Detailed derivations of the signal induced in a quarter coil are discussed
in Appendix~\ref{app:integrals}, the resonance frequency of a quarter coil in Appendix~\ref{app:resonance-frequency} and the effect of mirror currents in Appendix~\ref{app:mirror_currents}.

\section{Toroidal coils as beam position monitors}
\label{sec:coils-as-bpms}
\subsection{General considerations}
\label{sec:general-considerations}

Coils wound on a toroidal coil body are called \textit{Rogowski coils}, named after Walter Rogowski\,\cite{Rogowski1912}. Such devices are used as AC current transformers and as beam current monitors, taking advantage of the fact that
due to Ampere's law, the induced voltage in the coil is \textit{independent} of the position of the current-carrying wire passing through it\,\cite{doi:10.1063/1.1135946,doi:10.1063/1.1136119,doi:10.1063/1.4916094,samimi:2015}. Rogowski coils are ideally suited for the application as BPMs because the magnetic flux lines of a current-carrying straight wire form concentric circles around the wire and penetrate the cross-sectional area of the windings of the Rogowski coil at right angles, resulting in an optimal induced potential difference between the ends of the coil. 

\subsection{Position response of differential coil setup}
\label{sec:position-response-diff-sensor}

\begin{figure}[tb]
\begin{center}
\includegraphics[width=1\columnwidth]{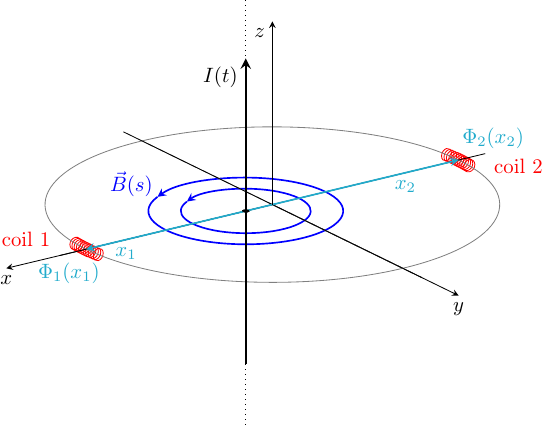}
	\caption{\label{fig:position-response} Beam position response of two identical induction coils (1 and 2). The $z$ axis points along the direction of the beam current $I(t)$. The change of the induced magnetic flux $\Phi_1$ and $\Phi_2$, probed by the two coils, is sensitive to the displacement of the beam from the center $\Delta x = (x_2 - x_1)/2$.}
\end{center}
\end{figure}

The magnetic difference signal of two identical induction coil sensors is a measure of the position of the beam between them, as derived in this section. The magnetic flux density generated by the beam current $I(t)$ can be written as
\begin{equation}
 \vec B(t) = \frac{\mu_0 I(t)}{2\pi\,  s} \vec e_\textbf{t} \,,
\end{equation}
where $\vec e_\textbf{t}$ is a unit vector along the circumference,
$\mu_0$ is the vacuum magnetic permeability and $s$ is
the radial distance from the beam center. As shown in
Fig.\,\ref{fig:position-response}, the beam is perpendicular to the plane
defined by the two coils. 

The time derivative of the induced flux in two identical short induction coils, 1 and 2, located on a circle of diameter $d$ is given by 
\begin{equation}
\dot{\Phi}_{1,2} = -\frac{U_{1,2}}{N_w} = \dot{B}_{1,2} \cdot S = \frac{\mu_0\dot{I}}{2\pi} \frac{S}{x_{1,2}}\,,
\end{equation}
where $S$ denotes the cross-sectional area of the coils. The ratio of induced voltage difference to voltage sum in the two coils is therefore given by 
\begin{equation}
\begin{split}
     \frac{\Delta U}{\Sigma U} = \frac{U_1 - U_2}{U_1 + U_2}  = \frac{x_2 - x_1}{x_2 + x_1} = \frac{2\Delta x}{d} 
\end{split}     \,.
\label{eq:derivation-linear-position-dependence}
\end{equation}
The above considerations also apply to rotationally symmetric charge distributions of the beam, as discussed in more detail in Sec.\,\ref{sec:flux-in-a-quarter-coil}.

\begin{figure}[tb]
\begin{center}
    \includegraphics[width=1\columnwidth]{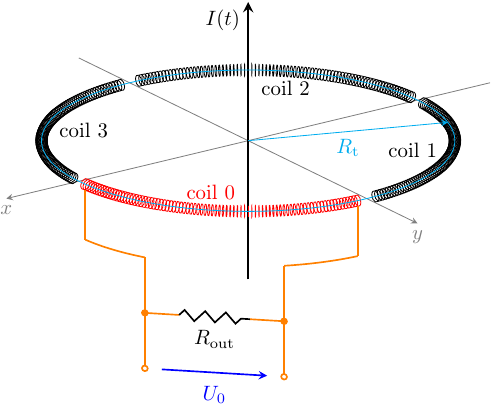}
\end{center}
	\begin{center}
		\caption{\label{fig:frequency-response} Toroidal coil of radius $R_\text{t}$, segmented into four quadrants ($M=0,1,2,3$). The (beam) current $I(t)$ passes through the coil along the $z$ axis. The induced voltage $U_0$ in one quarter is measured by a device with an input impedance of $R_\text{out}$ (see also equivalent circuit diagram in Fig.\,\ref{fig:equivalent-circuit}).}
	\end{center}
\end{figure}

The above derivation illustrates that a suitably segmented toroidal coil may serve as a BPM. Due to symmetry considerations, the induced voltage difference is an odd function of the beam displacement, thus yielding a linear relationship for small displacements, as illustrated in Eq.\,(\ref{eq:derivation-linear-position-dependence}). A toroidal coil segmented into four elements, as indicated in Fig.\,\ref{fig:frequency-response}, allows one to simultaneously determine the beam position along two orthogonal coordinate axes. This specific segmentation will be discussed further in this publication.

\subsection{Evaluation of the flux in a quarter coil}
\label{sec:flux-in-a-quarter-coil}

For the calculation of the flux encircled by the windings on the torus, we start from the vector potential $\vec A$, which is linked to the beam current density $\vec j$ via Poisson’s equation. Using cylindrical coordinates, $\rho, \varphi$ and $z$, one finds 
\begin{equation}
    \Delta \vec A(\rho,\varphi,z) = -\mu_0 \cdot \vec j (\rho, \varphi, z)\,,
\end{equation}
when displacement currents are neglected. If the beam current has only a
component in $z$-direction, as shown in Fig.\,\ref{fig:position-response}, the
same holds for the vector potential $\vec A$, and we can regard the problem as
two-dimensional. Furthermore, we consider here beams with rotationally
symmetric current distributions that are much smaller than the coil diameter
$d = 2R_\text{t}$, so that no beam particles intercept the Rogowski coil. Such a current distribution generates a field in its outside region that bears no information about the radial current distribution, and is therefore equivalent to a pencil beam. In this case, the remaining component of the vector potential $A_z$ is a harmonic function outside the beam, i.e., $\Delta  A_z =0$, which is given by the logarithm of the distance to the beam. Thus, in cylindrical coordinates we arrive at the ansatz
\begin{equation}
	A_z(\rho, \varphi) = -\frac{\mu_0 I}{2\pi}  \ln \left(
        \sqrt{(\rho\cos\varphi -x)^2 + (\rho\sin \varphi -y)^2}   \right)\,.
	\label{eq:Az-of-rho-and-phi}
\end{equation}
In order to find the induced flux, an integration of $A_z$ along the wire path $\vec \ell$ has to be performed.

The calculation of the induced magnetic fluxes $\Phi_M$ in the four quadrants $M=0,1,2,3$, discussed in Appendix\,\ref{app:integrals}, yields a power expansion in terms of the beam displacements $x$ and $y$. The corresponding induced voltages for any Fourier component of the periodic beam current in the storage ring are given by differentiation, which is conveniently carried out in the frequency domain by using $\omega = 2 \pi f$. 

Making use of Eq.\,(\ref{eq:f-of-omega}) considering the amplitudes
\begin{equation}
U_M = F(\omega) U_M^{\text{ind}} = -F(\omega) \dot{\Phi}_M = -F(\omega) \omega \Phi_M \, ,
\label{eq:uhat-of-omega}
\end{equation}
where the frequency response $F(\omega)$  
is included because the quarter coil is operated in a resonance regime.
Using Eq.~(\ref{eq:flux-compact}) one obtains for the induced voltage in a quadrant $M$
\begin{eqnarray}
\lefteqn{U_M(x,y,\omega) = F(\omega) U_M^{\text{ind}}(x,y) }\nonumber \\
 &=& \hat U(\omega)
 \left( D_0(b) + \sum\displaylimits_{m=1}^{\infty} D_m(b) \cdot  E_{m,M}(x,y) \right) \, , \label{eq:induced-voltage-quarter-0}
 \\
 &&\text{where} \quad  \hat U(\omega) = \mu_0 a N_{\text{w}} \, \omega F(\omega) I \, ,  \nonumber 
\end{eqnarray}
$N_\text{w}$ is the number of windings and
$b$ denotes the ratio of small radius $a$ to large radius $R_\text{t}$ of the toroid (see Table\,\ref{table:basic-parameters} and Fig.~\ref{fig:position-response-appendix}). The geometric functions $D_m(b)$ are defined in Eq.\,(\ref{eq:D0-and-Dm}), listed in Table\,\ref{table:Dm-cs-and-ds}, and analytical expressions for the $E_{m,M}(x,y)$ are given in Table\,\ref{table:Ems} in Appendix\,\ref{app:integrals}. For vanishing beam displacements $ E_{m,M}(0,0) = 0$, the remaining $D_0(b)$ describes the common voltage induced in each of the quadrants. 
The frequency response $F(\omega)$ is discussed in Sec.\,\ref{sec:frequency-response} and App.~\ref{app:resonance-frequency}.

The symmetries that apply to the induced flux in the different quadrants are passed on to the induced voltages of the quadrants [see Eq.\,(\ref{eq:symmetry-relation-for-induced-flux})]. Given the induced voltage $U_0(x,y)$ from Eq.\,(\ref{eq:induced-voltage-quarter-0}),  one can write for the induced voltages in the other quadrants,
\begin{equation}
\begin {split}
U_1(x,y) & = U_0(-x,y) \, , \\
U_2(x,y) & = U_0(-x,-y) \, ,\\
U_3(x,y) & = U_0(x,-y) \,.
\end{split}
\label{eq:symmetry-relation-for-induced-voltages}
\end{equation}
Therefore, assuming identical segments, the sum over the voltages of all quadrants is independent of the beam displacements, 
\begin{equation}
U^\Sigma = 
\label{eq:sum-of-induced-voltages}
\sum_{M=0}^3 U_M = 4 \, \hat{U}D_0(b)\,.
\end{equation}
This property makes it possible to monitor the current of a bunched beam with
an \textit{unsegmented} Rogowski coil.

Small beam displacements can adequately be described by the linear term in Eq.\,(\ref{eq:induced-voltage-quarter-0}), corresponding to $m = 1$ in the sum. Thus, to lowest order
\begin{eqnarray}
\lefteqn{U_0  \approx \hat{U}(\omega) \cdot \left[ D_0(b) + D_1(b) \cdot  E_{1,0}(x,y) \right]}\label{eq:U0-to-lowest-order} \\
    & =& \hat{U}(\omega) \cdot \Big[\frac{1-\sqrt{1-b^2}}{b} \, \frac{2\Delta \theta}{\pi} \nonumber \\
    && + \frac{2}{b \pi} \left( \frac{\displaystyle 1}{\displaystyle \sqrt{1-b^2}}-1   \right)         \frac{x+y}{R_\text{t}}  \, (\cos(\theta) - \sin(\theta))                       \Big]  \nonumber 
\end{eqnarray}
The linear term in the displacements $x,y$ is the basis for the operation of opposed quadrants or halves as BPMs, if the corresponding signals are subtracted.

\begin{figure}[htb]
\begin{center}
  \includegraphics[width=1\columnwidth]{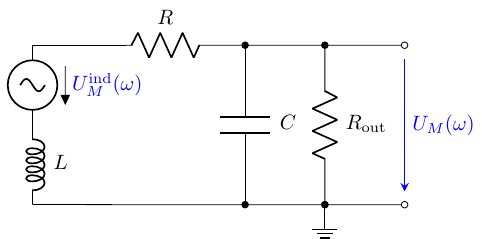}
\end{center}
\begin{center}
		\caption{\label{fig:equivalent-circuit} Equivalent circuit diagram for one toroidal coil quadrant with inductance $L$, as shown in Fig.\,\ref{fig:frequency-response}. A voltage $U^{\text{ind}}_M(\omega)$   is induced due to the 
       inductive coupling of the beam inside the coil.
 The resistance of the quadrant coil is $R$, and its capacitance $C$ is mostly due to the connecting wires. $R_\text{out}$ represents the input impedance of the preamplifier.}
\end{center}
\end{figure}

\subsection{Frequency response}
\label{sec:frequency-response}

Before we discuss in Sec.\,\ref{sec:experimental-setup} the specifics of the design of a BPM based on a segmented toroidal Rogowski coil, the frequency response of such a system shall briefly be addressed. The basic setup of a toroidal coil that is split into four individual quadrants is shown in Fig.\,\ref{fig:frequency-response}. The equivalent circuit diagram for a single quadrant coil coupled to the beam circuit is depicted in Fig.\,\ref{fig:equivalent-circuit}.

The voltage $U_M$ due to the induction by the orbiting bunched beam is modified by the resistance $R$ of the coil wire, by the inductance $L$ of the coil, and by the capacitance $C$. The latter is mostly due to the connecting wires. The revolution frequency of the beam is low enough to treat these components as lumped elements, as shown in the equivalent circuit diagram in Fig.\,\ref{fig:equivalent-circuit}. The resistor $R_\text{out}$ describes the input impedance of the measuring device. 

These lumped elements, $C$, $L$, $R$, and $R_\text{out}$ constitute a resonant circuit. Its frequency response $F(\omega)$ with the values given in Table\,\ref{table:basic-parameters} is calculated in Appendix\,\ref{app:resonance-frequency}. This yields for a quarter coil an estimated resonant frequency of 
$ f_0 \approx \SI{5.88}{\mega \hertz}$\,.

One would ideally operate the coil quarters at resonance to take advantage of the
amplified voltage in order to improve the signal-to-noise ratio.
The disadvantage is that small shifts in the resonance frequency due to
external conditions (e.g.~temperature drifts) would lead to large changes in the amplifying
factors of the four coils. Due to manufacturing tolerances the four resonance
curves are not absolutely identical, which would therefore lead to changes in the
position measurement. For this reason the decision was made to operate them off-resonance.

The Rogowski coils were designed for operation 
at a frequency of \SI{3}{MHz}, i.e.~with four bunches orbiting in the storage ring at a revolution frequency of
\SI{750}{kHz}. The final experiment was carried out in a single-bunch mode, thus calibrations and measurements were performed at \SI{750}{kHz}, i.e.~far away from the resonance frequency.
The resonant frequency of each coil was adjusted to match the revolution frequency of the beam by placing an appropriate capacitor in parallel to the coil.

\begin{table}[tb]
\caption{\label{table:basic-parameters} Parameters of the  Rogowski coil design. The electrical parameters $C$, $L$, and $R$ are calculated in Appendix\,\ref{app:resonance-frequency}. Typical beam parameters for a momentum $p$ are listed. $e$ denotes the elementary charge.}
	\begin{ruledtabular}
		\begin{tabular}{lr}
			Parameter & Value \\ \hline
			Number of quadrants $M$ & $4$ \\
			Toroid large radius & $R_\text{t} = \SI{58.5}{\milli \meter}$ \\
			Toroid small radius & $a = \SI{6.0}{\milli \meter}$\\	
			Ratio parameter & $ b = a/R_\text{t} = 0.1026$ \\ 
			Wire diameter & $d_\text{w} = \SI{450}{\micro \meter}$ \\
			Windings per quadrant & $N_\text{w} = 132$ \\
             Angular coverage & $\Delta \theta =\SI{64}{\degree}$\\\hline
			Capacitance per quadrant (wiring) & $C \approx \SI{20.7}{\pico \farad}$ \\
			Inductance per quadrant & $L \approx \SI{41.4}{\micro\henry}$\\
			Resistance of quarter coil & $R \approx \SI{0.61}{\ohm}$\\
			Input impedance &   $R_\text{out} \approx \SI{500}{\kilo\ohm}$\\ 
			Resonance frequency (estimated) & $f_0 \approx \SI{5.88}{\mega \hertz}$ \\ \hline
			Beam momentum & $p = \SI{970}{MeV/c}$ \\
			Revolution frequency & $f_\text{rev} = \SI{750197.3}{\hertz}$\\
			Lorentz factor & $\beta = 0.459$  \\
			Number of stored particles & $N = \num{1e9}$\\
			Corresponding beam current & $I = e \, N \, f_\text{rev} = \SI{120.2}{\micro \ampere}$
		\end{tabular}
	\end{ruledtabular}
	
\end{table}

\section{Experimental realization}
\label{sec:experimental-setup}

\subsection{Physical parameters of the Rogowski BPM}
\label{sec:physical-parameters}

\begin{figure}[htb]
	\centering
    \includegraphics[width=\columnwidth]{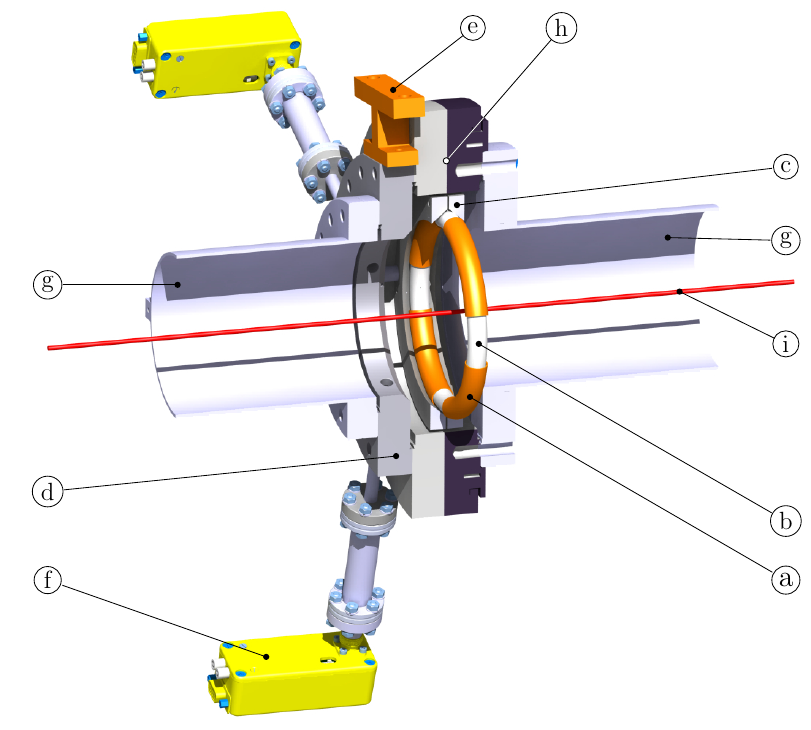}
	
	\caption{\label{fig:test-stand-1} Internal structure of the setup of the toroidal  coil support in the test stand, depicted in Fig.\,\ref{fig:test-stand-2}: 
		\protect\circled{a} Toroidal quarter coils on torus, 
		\protect\circled{b} coil supporting torus (PEEK), 
		\protect\circled{c} coil support rings (PEEK), 
		\protect\circled{d} DN 100/150 CF coil support flange 
		\protect\circled{e} fiducial mark, 
		\protect\circled{f} coaxial feedthrough with preamplifier, 
		\protect\circled{g} DN 150 CF beam line vacuum tubes, 
		\protect\circled{h} rotary flange, and 
		\protect\circled{i} current carrying wire.
	}
\end{figure}

The Rogowski BPM described here consists of four equal coil segments, as illustrated in Fig.\,\ref{fig:test-stand-1}, \circled{\small a}. The four corresponding voltage signals of the preamplifiers~\cite{ikpar2016:merzliakov} \circled{\small f} can be combined electronically to yield the differential signals of two sets of opposing half coils, which serve to simultaneously determine the beam displacement in $x$ and in $y$ direction.

PEEK plastic (Polyether ether ketone, material properties are given in  \url{https://pubchem.ncbi.nlm.nih.gov/compound/19864017})
 was chosen as the material for the coil body, because of its low cost, good machineability and low outgassing rates in vacuum. The earlier used  Vespel\textsuperscript{\textregistered} material (For material properties, see \url{https://www.dupont.com/products/vespel.html})  proved worse in terms of price, and showed higher outgassing rates, mostly due to absorbed water. The coil body features both a groove along its outer circumference and small radial bore holes for the returning quarter coil wires and their fixation, respectively.

The inner and outer diameter of the toroid onto which the Rogowski coils are  wound amount to \SI{105}{mm} and \SI{129}{mm}, respectively, sufficiently large to avoid any obstruction of the beam inside the circular beam tube diameter of \SI{150}{mm} in the COSY straight sections. The Kapton-insulated copper wire for the sensor coils has a diameter of \SI{450}{\micro \meter}. 
The number of windings ($N_\text{w} = 132$) determines the inductance of the quarter sensor coil and thus its resonant frequency, as discussed in Sec.\,\ref{sec:frequency-response}. The calculated ohmic resistance of the coil is $R\approx \SI{0.61}{\ohm}$. The inductance can be estimated from the textbook formula of a straight coil with the same length, which yields a value of $L  \approx \SI{41.4}{\micro \henry}$. 

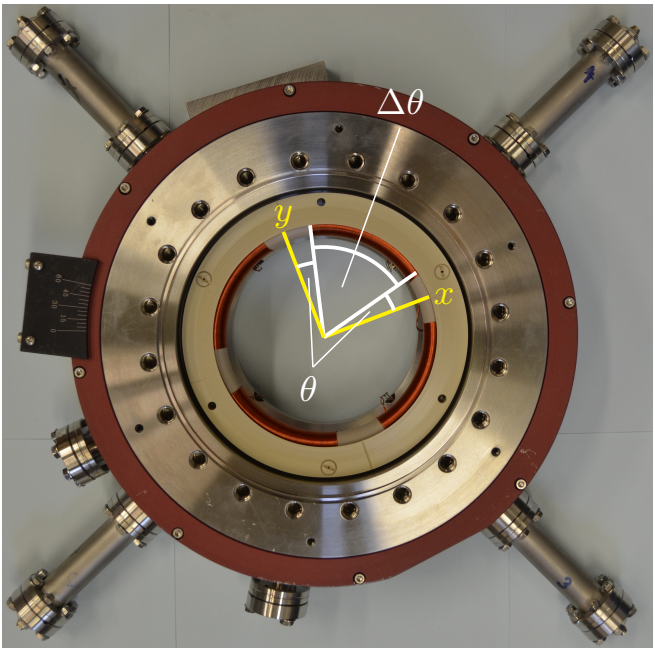
\begin{figure}[tb]
	\centering
	 \begin{tikzpicture}
      \draw(0,0) node{\includegraphics[angle=180,width=\columnwidth]{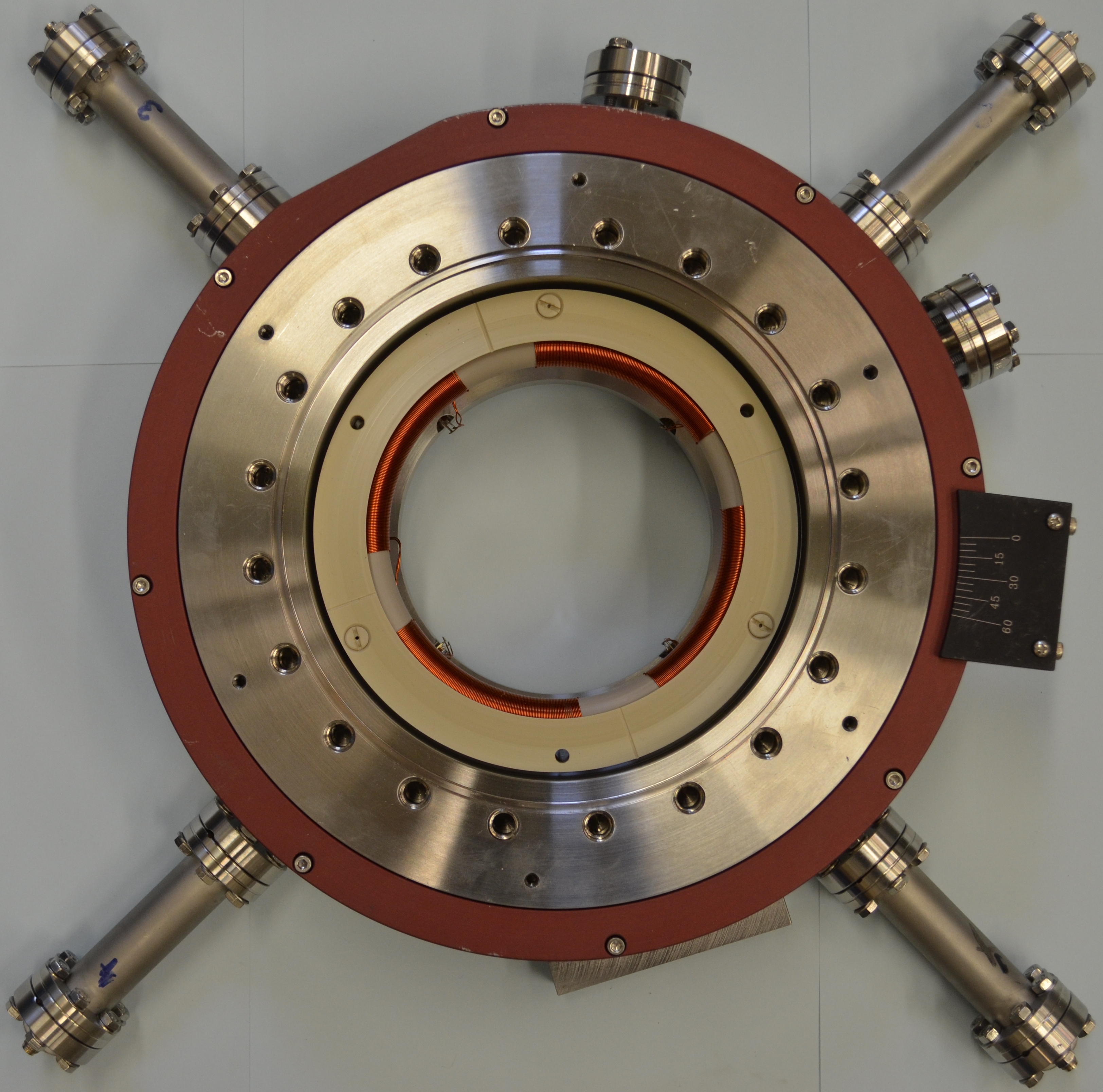}};
\begin{scope}[xshift=-0.5mm,yshift=-1.5mm]
   \draw[line width=0.5mm,yellow,rotate=21](0,0) -- (1.5,0) node[xshift=2mm,scale=1.5]{$x$};
   \draw[line width=0.5mm,white,rotate=21+14](0,0) -- (1.5,0);
   \draw[line width=0.5mm,white,rotate=21+90-14](0,0) -- (1.5,0);
   \draw[line width=0.5mm,yellow,rotate=21+90](0,0) -- (1.5,0) node[yshift=2mm,scale=1.5]{$y$};
   
   \draw[line width=0.5mm,white,rotate=21] (1,0) arc (0:14:1);
   \draw[line width=0.5mm,white,rotate=21+14] (1.2,0) arc (0:63:1.2);
   \draw[line width=0.5mm,white,rotate=21+90-14] (1,0) arc (0:14:1);

   \draw[line width=0.2mm,white,rotate=25] (0.5,0.5) --++ (1.6,1.6) node[scale=1.5,above,white]{$\Delta \theta$};

   \draw[line width=0.2mm,white,rotate=25] (-0.3,-0.3) -- (0.15,0.8) node[scale=1.5,below,yshift=-7.5mm,white]{$\theta$};
   \draw[line width=0.2mm,white,rotate=25] (-0.3,-0.3) -- (0.7,0.05) ;
\end{scope}
 \end{tikzpicture}
 
 \caption{\label{fig:foto-coil} Photograph of the Rogowki-BPM installed in
          the rotary flange as shown in Fig.\,\ref{fig:test-stand-1}. The coil covers an angle $\Delta \theta$.
          Note that $\Delta \theta + 2\theta=\SI{90}{\degree}$}
\end{figure}

To install the coil in the CF 160 flange, it is clamped between two PEEK rings held together by screws of the same material. Each coil features a twisted wire pair of 10 cm length followed by a 16 cm long SMA cable leading to an SMA feedthrough on a CF 16 flange. Fig.\,\ref{fig:foto-coil} shows a
photograph of the fully assembled BPM as it was used both on the test bench
and in the storage ring.

\begin{figure*}[htb]
	\includegraphics[width=0.75\textwidth]{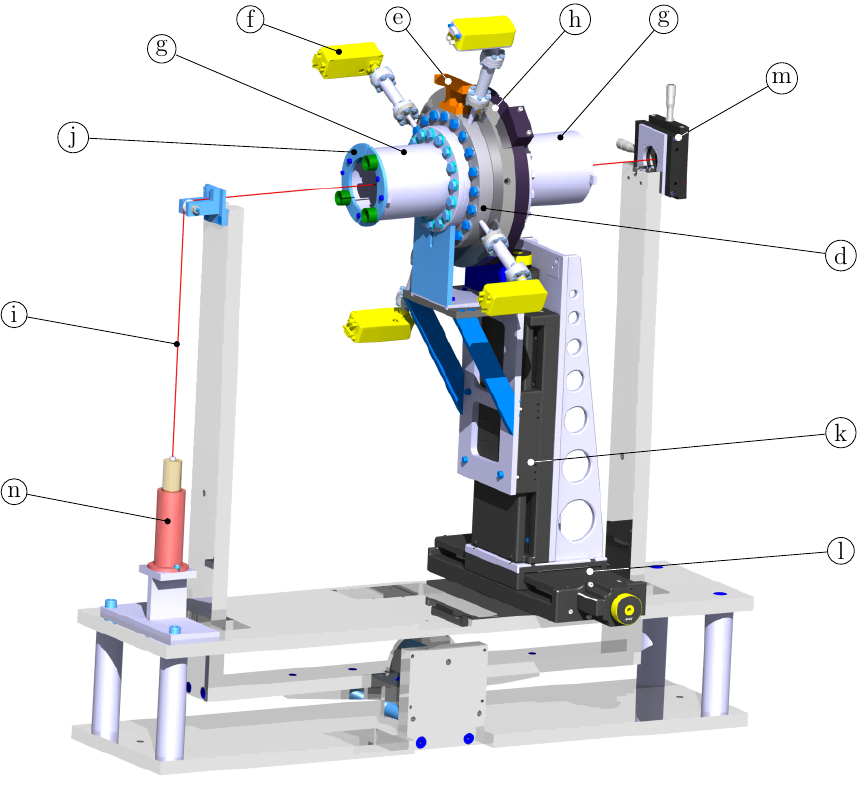}
	\caption{\label{fig:test-stand-2} Experimental setup to test and commission the Rogowski-type BPMs. The internal structure of the BPM is shown in Fig.\,\ref{fig:test-stand-1}.
		\protect\circled{d} DN 100/150 CF coil support flange, 
		\protect\circled{e} fiducial mark, 
		\protect\circled{f} coaxial feedthrough with preamplifier, 
		\protect\circled{g} DN 150 CF beam line vacuum tubes, 
		\protect\circled{h} rotary flange of the rf Wien filter,  
		\protect\circled{i} current carrying wire,
		\protect\circled{j} tool with knife egdes and fiducial marks (see Fig.\,\ref{fig:knife-edge}),
		\protect\circled{k} vertical ($y$) stepper-motor drive,
		\protect\circled{l} horizontal ($x$) stepper-motor drive,
		\protect\circled{m} manual $xy$-table for angular adjustment of the current-carrying wire, and 
		\protect\circled{n} weight in water bath to stretch the wire and to damp its oscillations.
	}
\end{figure*}

\begin{figure}[tb]
	\includegraphics[width=\columnwidth]{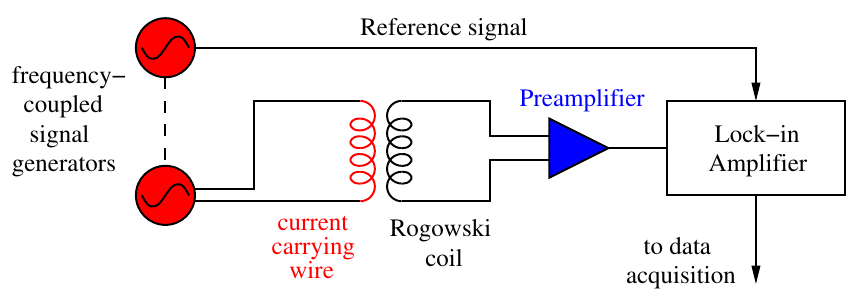}
	\caption{\label{fig:test-circuit} Electric circuit diagram for the measurements with the experimental test setup as shown in Fig.\,\ref{fig:test-stand-2}.
	}
\end{figure}

\begin{figure}[tb]
	\includegraphics[width=\columnwidth]{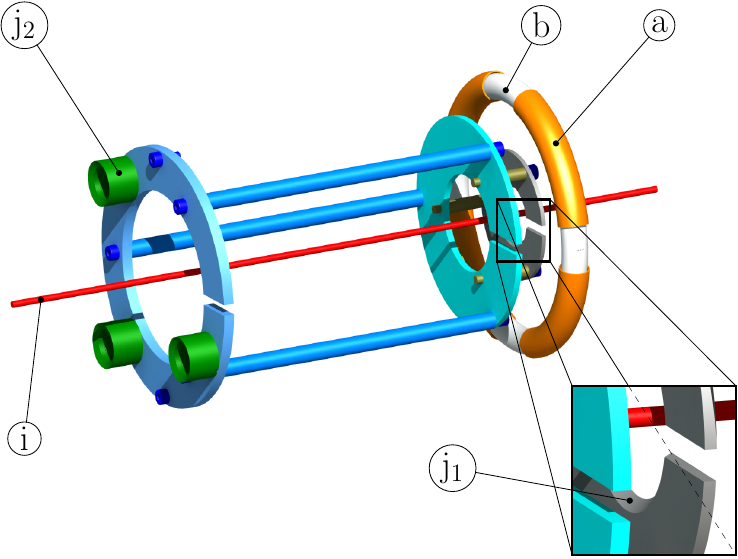}
	\caption{\label{fig:knife-edge} Knife edge tool with knife egdes
          \protect\circled{j$_1$} and fiducial marks
          \protect\circled{j$_2$}, quarter coils \protect\circled{a} wound on
          the torus \protect\circled{b}. The current carrying wire is denoted
          by \protect\circled{i}.
	}
\end{figure}

\subsection{Test stand for the Rogowski BPM}
\label{sec:experimental-setup-test-stand}

The laboratory test stand, shown in Fig.\,\ref{fig:test-stand-2},  has been
developed for the purpose of conditioning and calibrating Rogowski BPMs prior
to the installation at COSY. The BPM is fixed on top of a stepper motor-driven $xy$-table
\circled{\small k} \circled{\small l} in order to displace it laterally with
respect to a current carrying tin-coated (corrosion-free) copper wire \circled{\small i}, which
mimics the particle beam in COSY. A
sinusoidal current from a signal generator (Keysight 33522B Waveform Generator, Keysight Technologies, \url{www.keysight.com}) of about \SI{100}{\micro A} in amplitude represents the dominant Fourier component of the COSY beam current at the revolution frequency of \SI{750}{\kilo \hertz}. The wire of about \SI{1}{m} length is fixed on one side in a $xy$-table \circled{\small m} for angular adjustment  and on the other side it is guided by a roll and pulled by a weight in a water bath \circled{\small n} to assure a constant tension on the wire without creating vibrations. It is terminated by a \SI{50}{\ohm} resistor.

Tubes \circled{\small g} on both sides of the flange are added to mimic the electrical surroundings of the BPM in COSY. The stepper motors \circled{\small k} \circled{\small l} (LIMES 150-100-MiSM-IMS produced by OWIS GmbH, \url{www.owis.eu}) move the coil in the $xy$-plane during a calibration measurement. They have a
maximum travel range of \SI{100}{mm}, a load capacity of \SI{60}{kg} and a positioning error of $<\SI{10}{\micro m}$. The repeatability of a specific position is better than \SI{500}{nm}, the resolution of \SI{50}{nm} is limited by an encoder for each axis.

The analog signals from the coils are fed into custom-made
preamplifiers\cite{Bohme:2018sjy} 
\circled{\small f} with gain factors of about 18, measured at \SI{750}{\kilo
  \hertz}. The amplified signals are analyzed by lock-in
amplifiers (Zurich Instruments HF2LI Lock-in Amplifier 50 MHz, 210
  MSa/s \url{www.zhinst.com}) using the TTL signals from a second signal
generator of the same type as reference (Fig.\,\ref{fig:test-circuit}). This was necessary to avoid cross talk between the original signal and the reference signal when using two channels of the same device. Both generators were coupled to each other to assure that the generated frequencies were identical.

The analog-to-digital converters feature a resolution of 14 bit. A measurement bandwidth of \SI{6.81}{\hertz} with a fourth-order low-pass filter is typically used for signal processing. This entails a time constant of about \SI{10}{ms}. The demodulator needs a 10 times larger time to reach 99\% of the output level, which means that we are dealing with a minimum integration time of about \SI{100}{ms}.

\subsection{Calibration of the Rogowski BPM}
\label{sec:calibration}
For the relative calibration the device is moved relative to the current
carrying wire using the stepper motors \circled{\small k} \circled{\small l}
while the response of the quarter coils is processed by the readout
electronics. A map is recorded covering the central part of the space inside
the device (see Sec.~\ref{sec:calib-measure}). 

The absolute calibration is done using the insertion shown in Fig.\,\ref{fig:knife-edge}. This insertion was manufactured to determine the exact position of the wire during the calibration measurements. It consists of a structure which positions a knife edge \circled{\small j1} in the plane of the Rogowski torus \circled{\small b} and fiducial marks
\circled{\small j2} to be used for a laser tracking system. The exact
geometry was surveyed by a measuring machine (ZEISS UPMC850, Carl Zeiss QEC GmbH \url{www.zeiss.de}) to an
accuracy of $< \SI{1}{\micro \meter}$. This insertion can be installed at any
time making use of the slits in the discs without changing the position of the current wire. The wire
\circled{\small i} is
put at a voltage potential and a programmed routine moves the Rogowski BPM
with the insertion until the wire touches the knife edge. Using the
known geometry of the insert, the readings of the $xy$-tables and a laser
tracker (OMNITRAC 2, Automated Precision Inc. \url{www.apimetrology.com}), the absolute position of the measured calibration maps can be connected to
the fiducial mark\,\circled{\small e} located on
top of the Rogowski BPM (see Fig.\,\ref{fig:test-stand-2}) with an accuracy of
$< \SI{40}{\micro \meter}$. 
After installation of the BPM in COSY the position
of the fiducial mark is determined within the COSY lattice
using a laser tracker.
Thus the absolute position of the particle
beam within the COSY lattice can be determined 
with an accuracy defined by the accuracy of the measurement of the laser tracker, which is about $50$ to
\SI{100}{\micro \meter}. 

\subsection{Vacuum compatibility}
\label{sec:vacuum-compatibility}

Besides the accuracy of the position measurement, the compatibility with the vacuum standards of the accelerator is crucial. All components of the Rogowski coil were cleaned and baked before assembly. In order to test the vacuum compatibility before installation in the accelerator, a vacuum test stand was set up, which consists of a bakeable chamber that is pumped by a \SI{250}{l/s} turbo-molecular pump (comparable to the pumping speed at the installation point in the accelerator). A quadrupole mass analyzer was used to determine the composition of the rest gas in the chamber.

The use of  PEEK as material for the coil body and the Kapton coated wire for the coils yielded small outgassing rates. 
The pressures reached after baking at \SI{120}{\degreeCelsius} for a week are $< \SI{5e-9}{\milli \bar}$. The mass spectrum before baking showed a dominant
water peak. After baking the water is removed from the structure and the wire. To minimize water accumulation
on the BPM during installation in the ring, the time it was exposed to air was minimized.

Attention was paid to the temperature during the baking process, as this takes place after the calibration of the coil (which was carried out in air, see Sec.\,\ref{sec:calibration} and Fig.\,\ref{fig:test-stand-2}). The reason for this is that both the Kapton-coated wire and the PEEK plastic adsorb water in air. To ensure that baking did not alter the calibration values, several test calibrations were carried out before and after baking, which showed that there was no difference in the calibration values within previously mentioned accuracy. 

\section{Investigations using the test stand}
\label{sec:investigations-test-stand}

\subsection{Impedance measurement of a quarter coil}
\label{sec:impedance-meas}

\begin{figure}[tb]
\includegraphics[width = 0.48\textwidth]{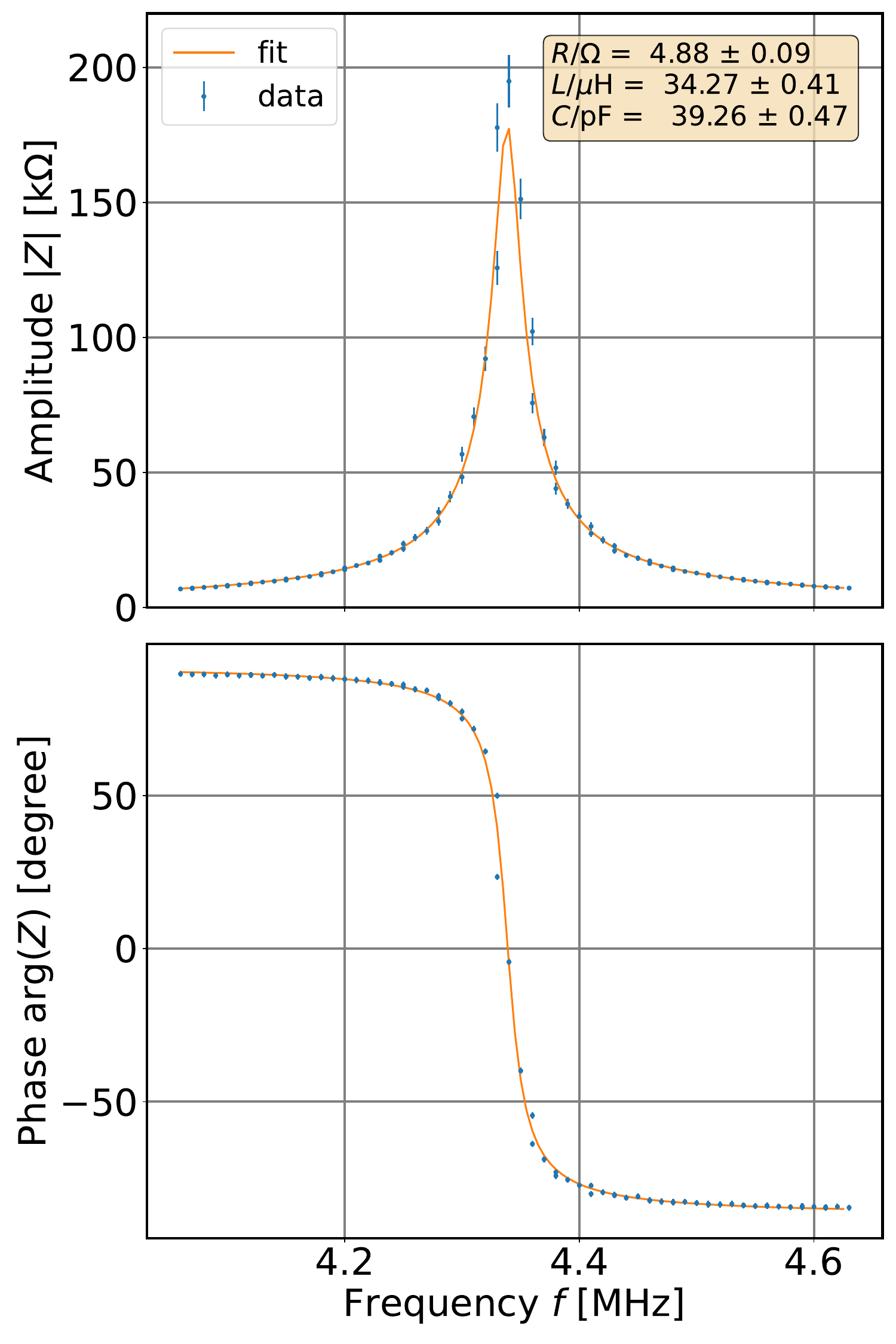}
\caption{\label{fig:impedance-coil}Combined fit to the amplitude and phase
  data from the network analyzer according to Eq.\,\ref{eq:impedance-Z}. 
}
\end{figure}

In order to characterize one quarter coil according to the circuit diagram of
Fig.\,\ref{fig:equivalent-circuit} an impedance measurement was performed
using a network analyzer (Siglent SVA 1032X, SIGLENT Technologies
  \url{www.siglenteu.com}). The influence of the connection cable outside the vacuum vessel 
was taken into account in the calibration, so that the results, shown in Fig.\,\ref{fig:impedance-coil}
represent the coil with the connection wires up to the vacuum flange
(see Fig.\,\ref{fig:test-stand-1}). The preamplifier was not used for this
measurement.

The impedance of the coil can be expressed by
\begin{equation}
Z(\omega) = \left( i \omega C + \frac{1}{R + i \omega L} \right)^{-1}  \, ,
\label{eq:impedance-Z}
\end{equation}
where $R$, $L$ and $C$ are the respective properties of the circuit, shown in
Fig.\,\ref{fig:equivalent-circuit}.
The amplitude and phase of Eq.\,(\ref{eq:impedance-Z}) were used in a combined
fit to determine these properties (Fig.\,\ref{fig:impedance-coil}).  
To improve the fit an additional offset of the
phase, was introduced as a parameter. The value amounts to $\SI{-2.9\pm0.1}{}$ degrees.
For the amplitude $|Z|$ relative error of 5\% was assumed, for the phase arg($Z$) the
assumed error is 1 degree.
These errors were estimated from the RMS of several measurements at the
same frequency.
The overall $\chi^2$ is 540. The amplitude contributes with 85 and the
phase plot with 455 to the $\chi^2$ value.
The number of degrees of freedom is 160 (data points) - 4 (parameters) = 156.
A possible correlation between amplitude and phase data is not taken
into account. The resulting properties are the ohmic resistance
$R= \SI{4.88 \pm 0.09}{\ohm}$, the inductance $L= \SI{34.27 \pm
  0.41}{\micro\henry}$ and the capacitance $C= \SI{39.26 \pm
  0.47}{\pico\farad}$. 
 The DC resistance of the coil was measured to be about \SI{1}{\ohm}.
For a copper wire used in the experiment due to the skin effect one
expects an ohmic resistance increase by a factor of 4, roughly in agreement with the fit parameters.

\clearpage

\subsection{Measurement of resonant behavior of a quarter coil}
\label{sec:meas-res-behavior}
\begin{figure}[b!]
\includegraphics[width = 0.48\textwidth]{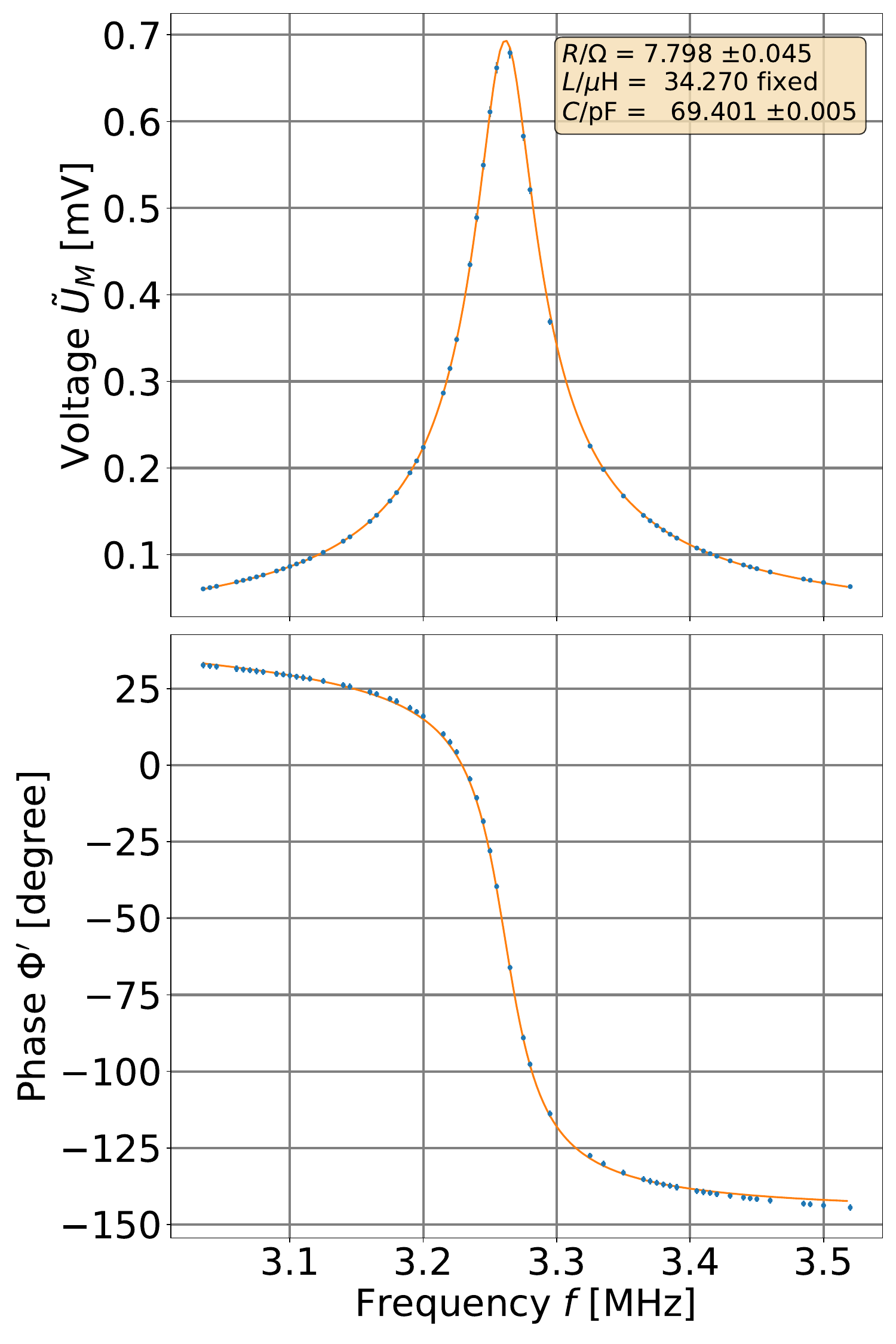}
\caption{\label{fig:toroid-coil-resonance-fits}Combined fit to the amplitude
  and phase data of one Rogowski coil excited by the current-carrying wire on the test stand. 
}
\end{figure}
The resonant behaviour of a quarter coil was measured on the test setup,
shown in Fig.\,\ref{fig:test-stand-2} using a wire to mimic the beam current. The
induced current in the four coils were amplified and analyzed in lock-in
amplifiers (see Sec.\,\ref{sec:experimental-setup-test-stand}). The result for
one quarter coil is shown in
Fig.\,\ref{fig:toroid-coil-resonance-fits}. 
The amplitude and phase are given by Eqs.~(\ref{eq:f-of-omega}) and (\ref{eq:phi-of-omega}).

To better fit the data, the heuristic
parameters $s_p, s_R$ and $\Phi_0$ were introduced, yielding
\begin{eqnarray}
 \tilde{U}_M(\omega) = (1&+&s_U \omega) F(R_\text{out},L,C,R',I,\omega)U_M^{\text{ind}}  \, ,\label{eq:voltage-Uout}\\
\mbox{where} \quad R' &=& R(1+s_R(\omega-\omega_0)) \, \\
 \phi'(\omega) &=& (1 + s_p \omega) \phi  - \phi_0 \, .
\label{eq:phase-Uout}
\end{eqnarray}
These additional parameters account for other
resonances present in the system and for the skin/proximity effect~\cite{terman1943radio}.
Due to the strong correlation between $C$ and $L$, the value of $L$ was fixed to the value from the impedance fit. An input impedance of the preamplifiers of $R_\text{out} = \SI{500}{k\ohm}$ was assumed.
The results are summarized in Table\,\ref{table:fit-resonance}.

\begin{table}[tb]
\caption{\label{table:fit-resonance} Parameters of the fit to the
          resonance data in Fig.\,\ref{fig:toroid-coil-resonance-fits}.}
	\begin{ruledtabular}
		\begin{tabular}{lr}
			Fit Parameter & Value \\ \hline
			$R$ & $\SI{7.798 \pm   0.045}{\ohm}$ \\
			$L$ fixed & $\SI{34.27}{\micro\henry}$ \\
                        $C$ & $\SI{69.401 \pm   0.005}{\pico\farad}$ \\
 			$R_\text{out}$ fixed & $\SI{500}{\kilo\ohm}$ \\
			$U_M^{\text{ind}}$ & $\SI{9.12\pm0.01}{\milli\volt}$ \\
 			$s_U$ & $\SI{8.0\pm0.2e-8}{\s^{-1}}$ \\
			$s_p$ & $\SI{-2.4\pm0.2e-8}{\s^{-1}}$ \\
            $s_R$ & $\SI{8.3\pm0.4e-7}{s^{-1}}$\\
 			$\phi_0$ & $\SI{152.5 \pm   0.2}{degree}$ \\
		\end{tabular}
	\end{ruledtabular}
	
\end{table}

For the $\tilde U_M$ a relative error of 1\% was assumed, for the phase the
assumed error is 1 degree. The $\chi^2$ is 52.4 and has a contribution of 16.3
from the amplitude and 36.1 from the phase. The number of degrees of freedom
is $2\cdot58 \text{(bins)} - 7 \text{(parameters)} =109$.

The equivalent circuit of Fig.\,\ref{fig:equivalent-circuit}, together with the additional parameters discussed above, describes the
data very well. The increased resistance $R$ and capacitance $C$ compared to
the impedance measurement are due to
the preamplifier. 

It was noticed that due to manufacturing tolerances  the resonance curves
slightly differed for the four coils. This resulted in a strong  frequency
dependence of the calibration. To reduce this effect variable capacitors were
installed, which shifted the resonance frequencies of all coils to a common
value. 

\subsection{Calibration measurement of the BPM}
\label{sec:calib-measure}
The calibration measurement was carried out as described in
Sec.\,\ref{sec:calibration}. A map of $21 \times 21 = 441$ points is measured
in the $xy$-plane in the range $\pm 10$ mm.
The four voltages $U_M$
of Eqs.\,(\ref{eq:induced-voltage-quarter-0},
\ref{eq:symmetry-relation-for-induced-voltages}) were recorded and processed as
follows. We define: 
\begin{equation}
\begin{split}
U^{\Delta x} &= U_0 - U_1 - U_2 + U_3 \, ,\\
U^{\Delta y} &= U_0 + U_1 - U_2 - U_3\, .\\
\end{split}
\end{equation}

\begin{figure}[b!]
\includegraphics[width = 0.48\textwidth]{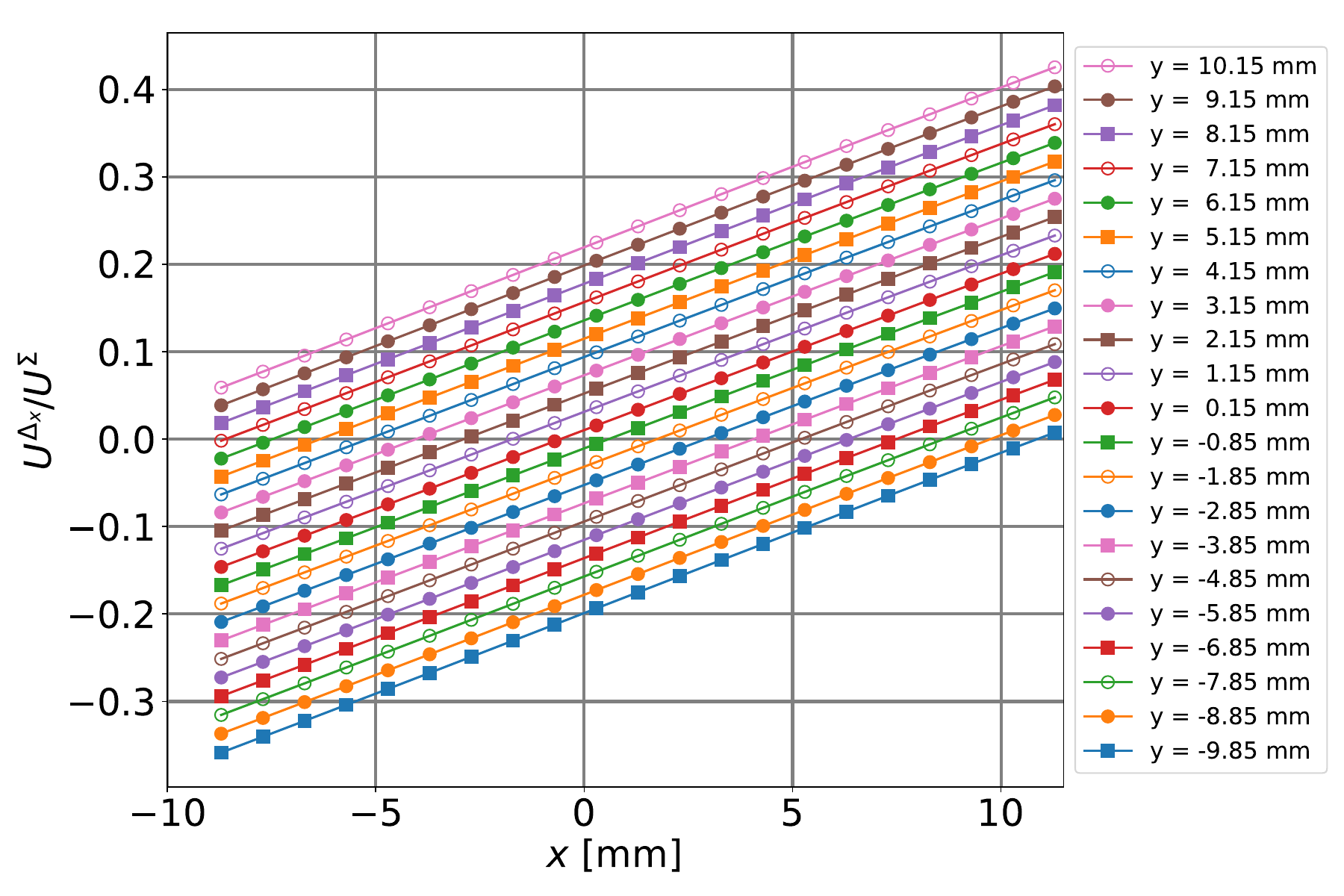}
\includegraphics[width = 0.48\textwidth]{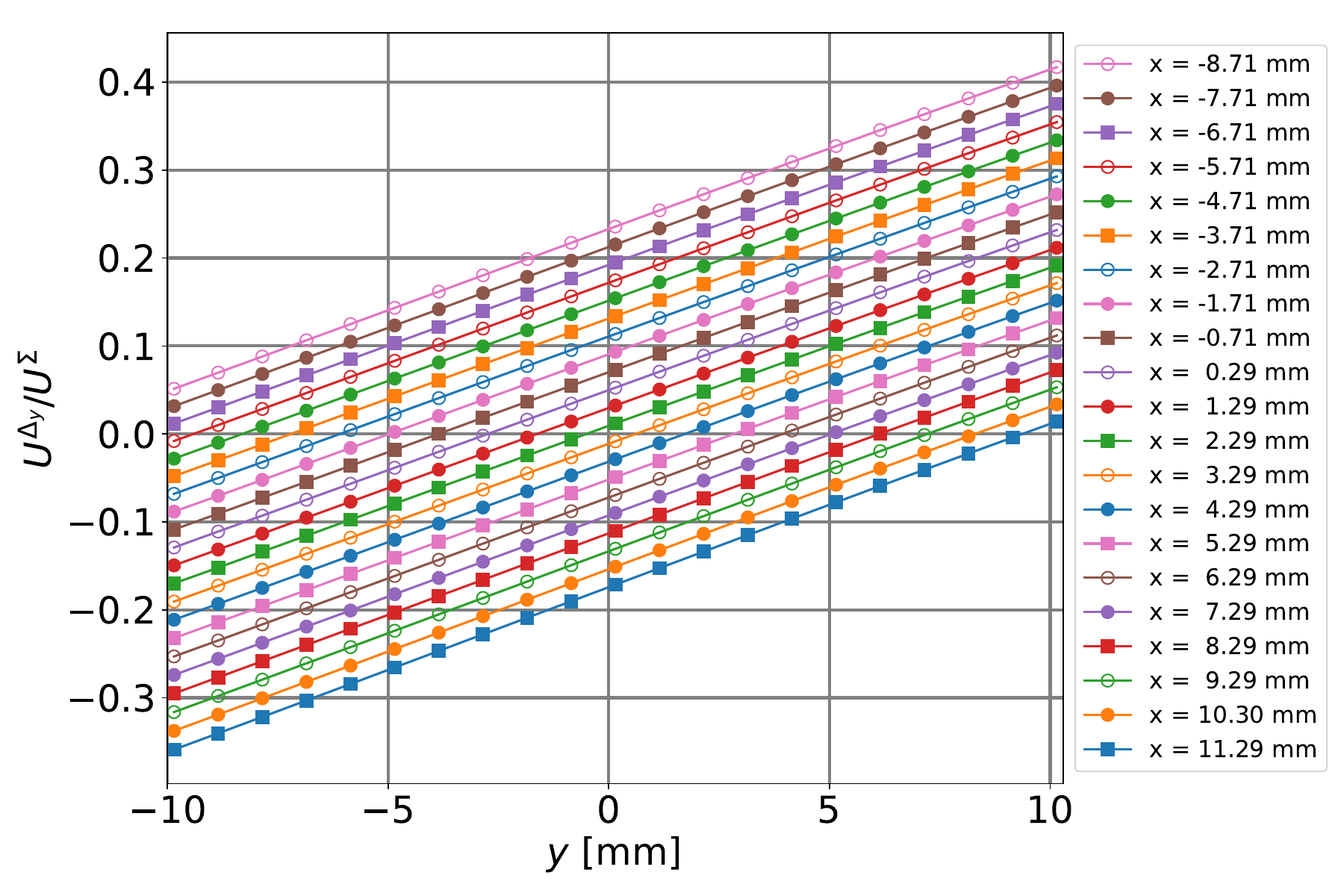}
\caption{\label{fig:Calibration}Calibration map with the measured
  $U^{\Delta x}/U^\Sigma$ and $U^{\Delta y}/U^\Sigma$ values depending on the
  respective $x$ and $y$ positions set with the stepping drives of the
  $xy$-table. For 
  better visibility the graphs are shifted by the respective $y$ and $x$ values.}
\end{figure}

Using $U^\Sigma$ from Eq.\,(\ref{eq:sum-of-induced-voltages}) the ratios
$U^{\Delta x}/U^\Sigma$ and $U^{\Delta y}/U^\Sigma$ are shown as a function of
$x$ and $y$, respectively,
in Fig.\,\ref{fig:Calibration}. For better visibility of the data the graphs
are shifted proportionally by the respective $y$ and $x$ values. The first order linear behavior clearly
dominates.
To  get a good description of the data by the theory additional parameters were
introduced: The fit includes terms up to 4th order in $m$ (see Appendix \ref{app:integrals}).
Despite of the clamping mechanism of the torus a perfect alignment
angle could not be guaranteed, therefore an angle $\alpha$ was added by
which the $xy$-coordinate system of the coils was rotated around the $z$-axis. The
coils themselves are not exactly equal, therefore small relative scaling coefficients
$(1+C_{sc})$ for the voltages for three of the coils with respect to the fourth were introduced.
From the RMS of repeated measurements the error on 
$U^{\Delta_{x,y}}/U^{\Sigma}$ was estimated to be $2 \cdot 10^{-4}$. With this error assumption a $\chi^2$-minimization
resulted 
in $\chi^2$ values of 19.5 and 7.5 for the
ratios $U^{\Delta_x}/U^\Sigma$ and $U^{\Delta_y}/U^\Sigma$,
respectively, for 441(measurement points) - 20(parameters) = 
421 degrees of freedom (ndf).
The low $\chi^2/\text{ndf}$ indicates that the errors
on $U^{\Delta_{x,y}}/U^{\Sigma}$ were probably overestimated.

\begin{figure}[htb]
\includegraphics[width = 0.5\textwidth]{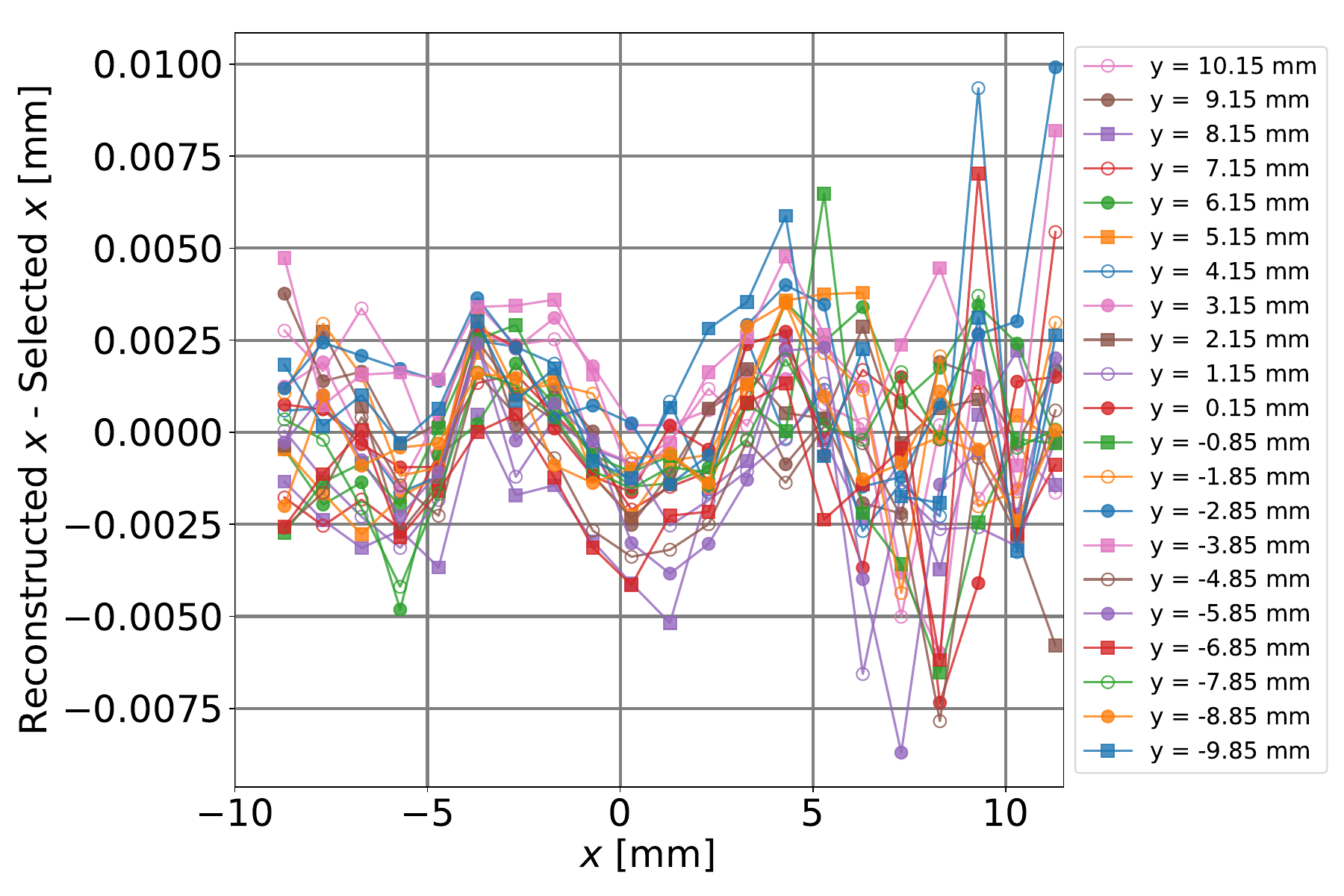}
\includegraphics[width = 0.5\textwidth]{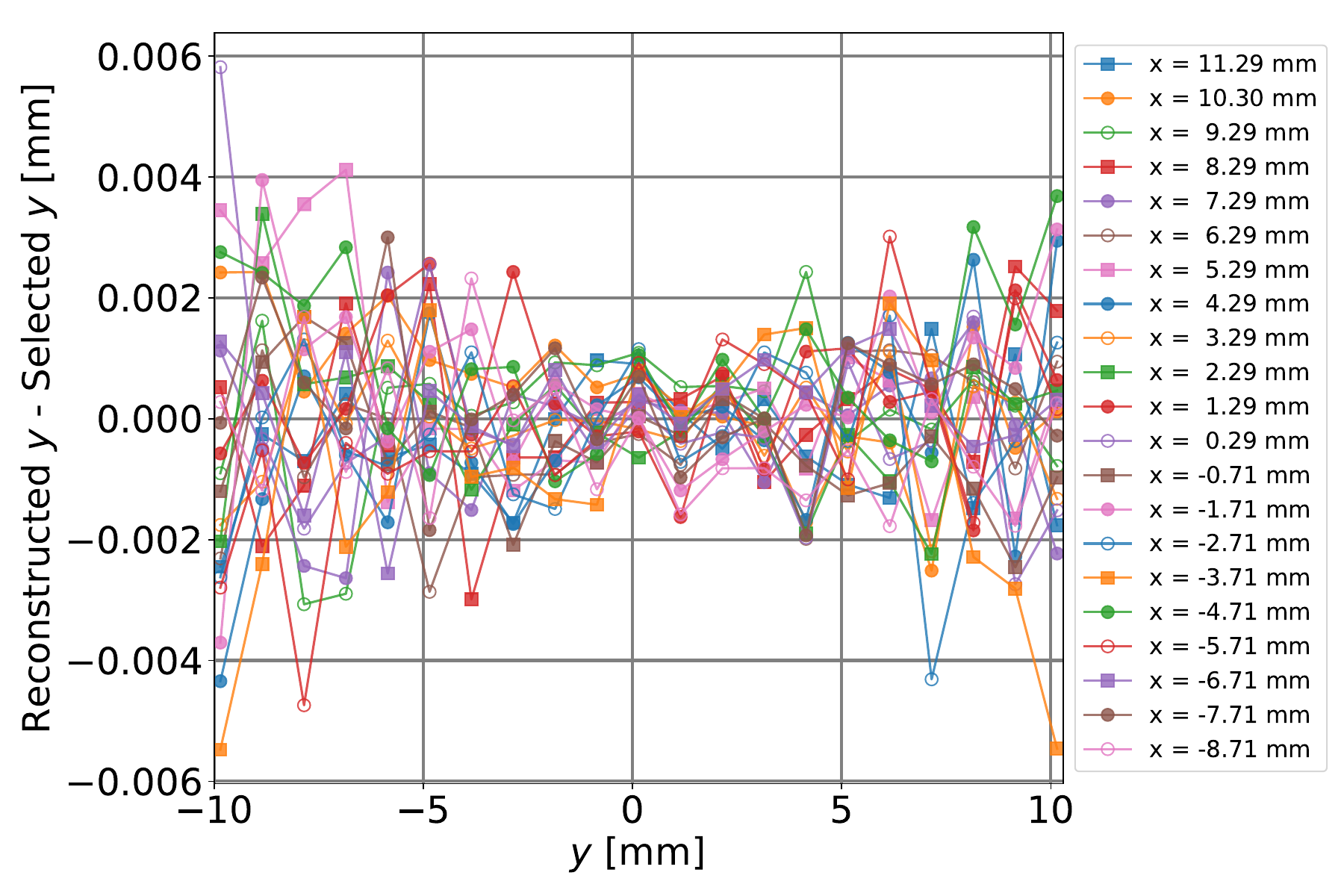}
\caption{\label{fig:Residuals} Residuals: reconstructed $x$ --
selected $x$ vs. $x$ (top)
	and reconstructed $y$ -- selected $y$  vs. $y$ (bottom).}
\end{figure}

From the fitted parameters to model $U^{\Delta_{x,y}}/U^{\Sigma}$,
one can now reconstruct $x$ and $y$ and compare it with
selected $x$ and $y$ values. This is shown in Fig.~\ref{fig:Residuals}.  The residuals (reconstructed - selected) are of the order of  few \si{\micro m}.

The linear coefficient $c_1$ of Fig.~\ref{fig:Calibration} can be derived from 
Eqs. \, (\ref{eq:symmetry-relation-for-induced-voltages} - \ref{eq:U0-to-lowest-order}) to:
\begin{eqnarray}
\label{eq:linear_coeff}
c_1 &=& \frac{D_1 E_1}{D_0}  \, \frac{1}{x+y} \nonumber \\
&=& \frac{1}{ R_\text{t} \sqrt{1-b^2}} \, 
 \frac{\cos(\theta) - \sin(\theta)}{\Delta\theta} \\
&\approx& \SI{0.0115}{mm}^{-1} \nonumber,
\end{eqnarray}
but the fitted values $c_1^\text{fit} \approx \SI{0.018}{mm}^{-1}$,
which could also be estimated directly from Fig.~\ref{fig:Calibration}, are
substantially higher.
This originates in
the surrounding of the coil. $c_1$ was calculated for an air coil in an
environment free of conducting material. Since the Rogowski coils are
installed inside the beam tube
(Fig.\,\ref{fig:test-stand-1}) secondary induction due to mirror currents
on the inner surface of the beam tube needs to be considered. This effect is
included in the coefficient (see Appendix~\ref{app:mirror_currents}, eq.~(\ref{eq:phi0_mc}))
\begin{eqnarray}
c'_1 &=& c_1 \left( 1 + \frac{b^2 R_r^2 }{2 u^2 b \pi D_1(b)} \right) \nonumber \\
&\approx& { c_1 \left(
1 +\frac{R_t^2}{2 u^2}  \right)} \approx \SI{0.015}{mm}^{-1}
\label{c1'-result}
\end{eqnarray}
with the inner radius of the beam pipe
$u = \SI{76.5}{mm}$.
This estimate is based on the simple geometry 
of a straight beam tube and explains qualitatively 
the difference between the fitted value and the
calculated value of an air coil in Eq.~\ref{eq:linear_coeff}.

\section{Installation at COSY}
\label{sec:installation-COSY}
\begin{figure}[tb]
\includegraphics[width = 0.48\textwidth]{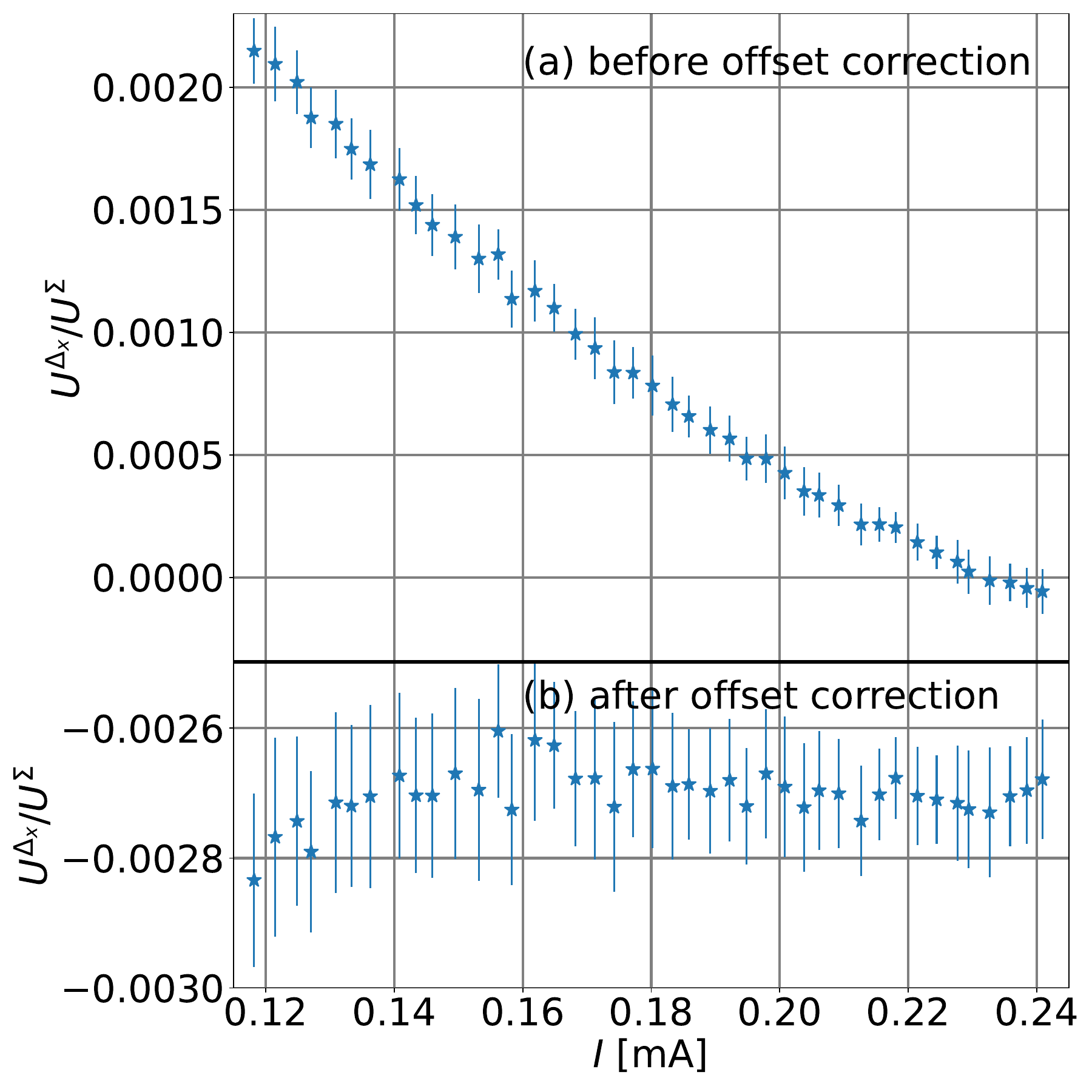}
\caption{\label{fig:COSY-correction-x} $U^{\Delta_{x}}/U^{\Sigma}$ before
  (a) and after (b) offset correction. Without the correction
  a drift depending on the beam current was observed in contrast to the
  measurements in the laboratory. The offset correction eliminates this dependence.}
\end{figure}

\begin{figure}[tb]
\includegraphics[width = 0.48\textwidth]{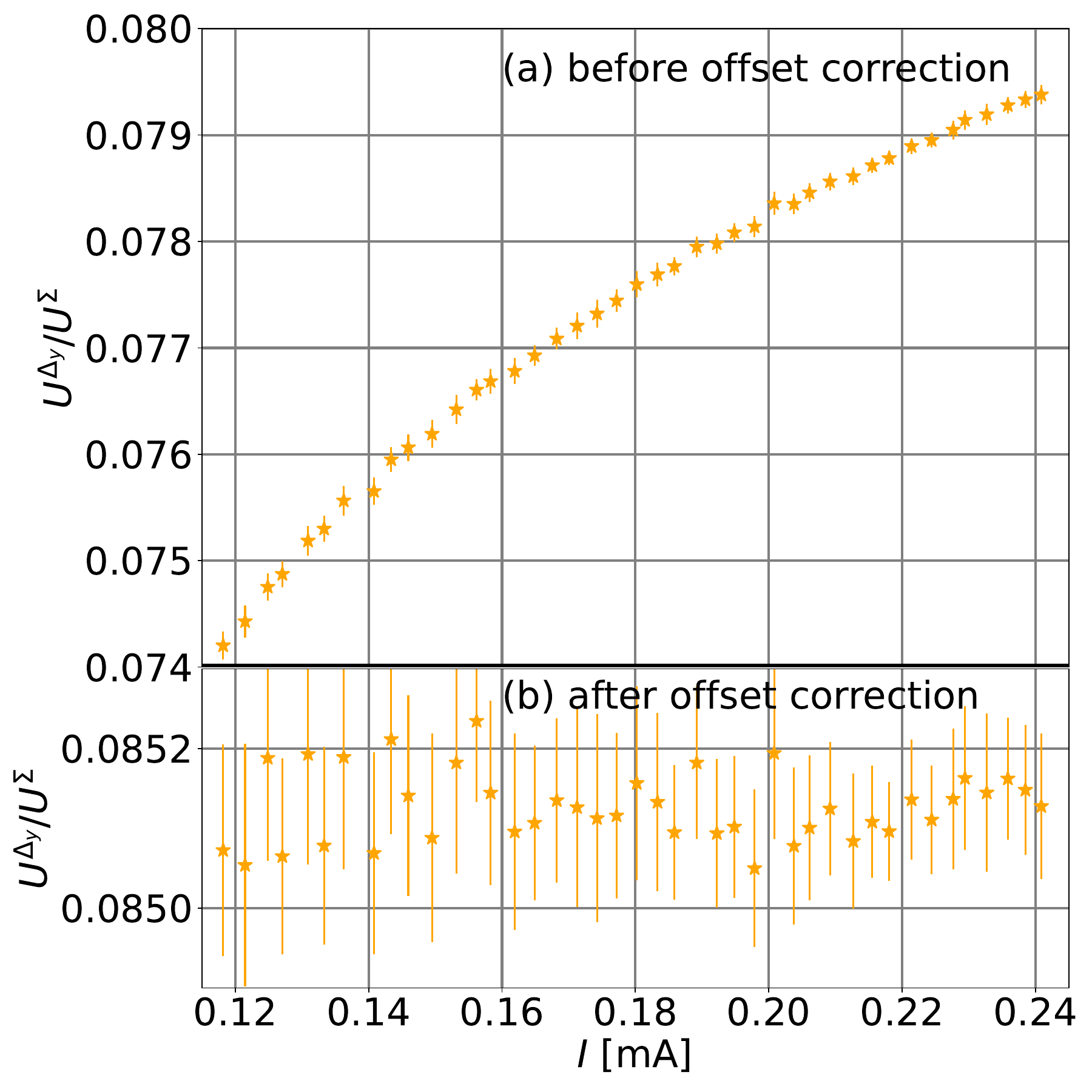}
\caption{\label{fig:COSY-correction-y} $U^{\Delta_{y}}/U^{\Sigma}$ before
  (a) and after (b) offset correction. Without the correction
  a drift depending on the beam current was observed in contrast to the
  measurements in the laboratory. The offset correction eliminates this dependence.}
\end{figure}

After the calibration (Sec.~\ref{sec:calib-measure}) and baking procedure
(Sec.~\ref{sec:vacuum-compatibility}) the coil was installed at the storage ring
COSY. It was
noticed that an additional offset voltage present at zero beam current
appeared. This offset disturbed the position measurements, so that the
position readings depended on the beam current,
as shown in Fig.~\ref{fig:COSY-correction-x}a and \ref{fig:COSY-correction-y}a.
The offset is due to stray fields of the COSY cavity
 which is located at a distance of \SI{9}{m} from
the Rogowski coils. The cavity is defining 
the COSY revolution frequency which is also used 
as a reference frequency of the lock-in amplifier.

\begin{figure}[htb]
\includegraphics[width = 0.48\textwidth]{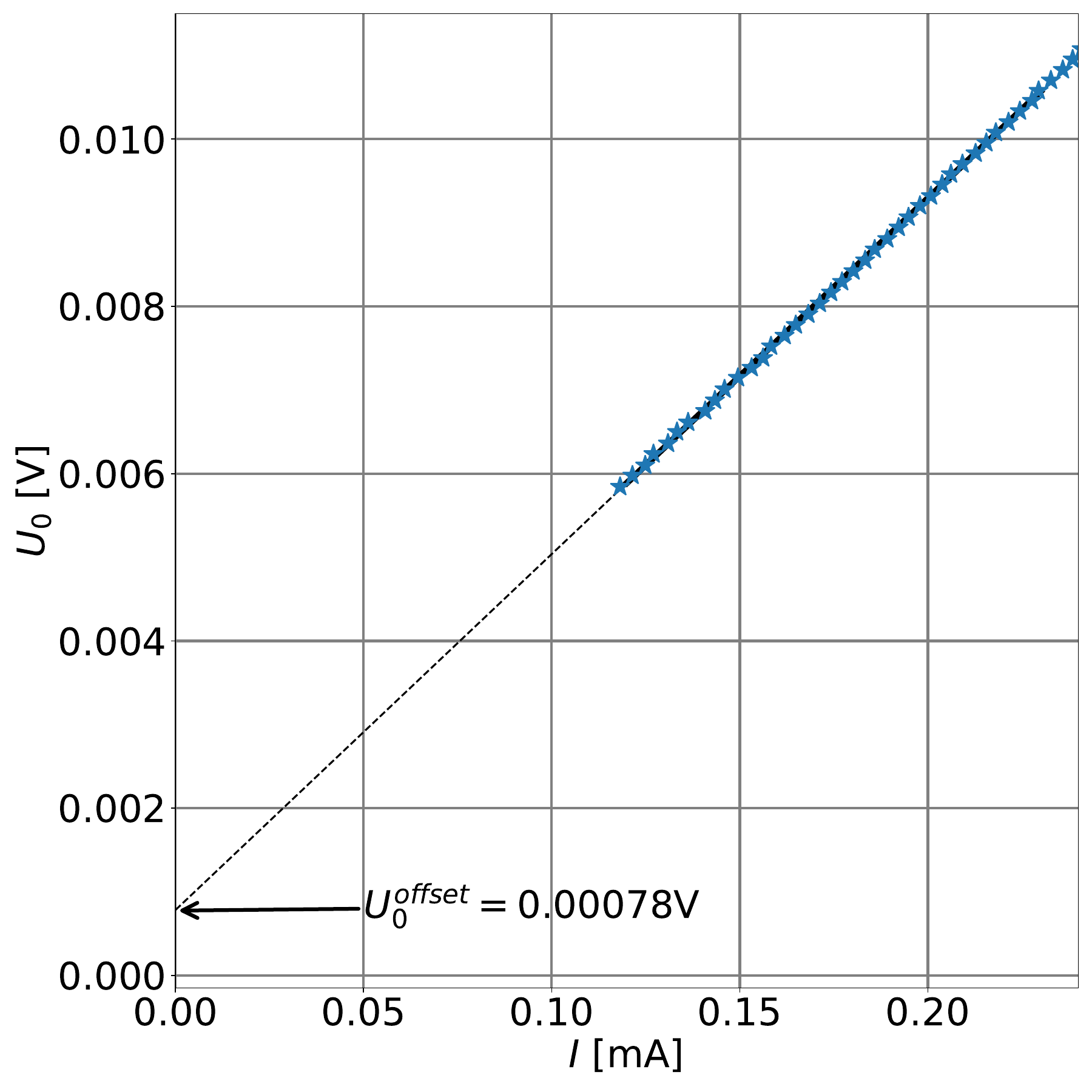}
\caption{\label{fig:COSY-offset-determination}
  Example of the induced voltage $U_0$ as a function of the COSY beam current $I$.
  The additional offset, due to stray fields, was determined by extrapolation to zero beam current.}
\end{figure}

The offset was determined by extrapolating the measured voltages to zero beam
current as can be seen for $U_0$ in Fig.~\ref{fig:COSY-offset-determination}.
The interception of the straight line fit with the vertical axis yields the
respective voltage offset corrections $U_0^{off} = \SI{775.4}{\micro V}$,
$U_1^{off} = \SI{749.5}{\micro V}$,
$U_2^{off} = \SI{718.2}{\micro V}$,
$U_3^{off} = \SI{706.5}{\micro V}$.
The subtraction of these offset values from the measured voltages resulted in beam
current independent position values as expected from the calibration
measurements in the laboratory (Fig.~\ref{fig:COSY-correction-x}b and
\ref{fig:COSY-correction-y}b). Note that a change of $10^{-4}$ in the voltage
ratio as shown in these figures corresponds 
to a change in position of $10^{-4}/c^{\mathrm{fit}}_{1} = 10^{-4}/0.018 \si{\micro m} \approx \SI{5.5}{\micro m}$.
It can be seen that the error on a single measurement of a \SI{1}{s} time
interval is also of a similar
amount, corresponding to a resolution of \SI{5}{\micro m}.

\begin{figure}[tb]
\includegraphics[width = 0.48\textwidth]{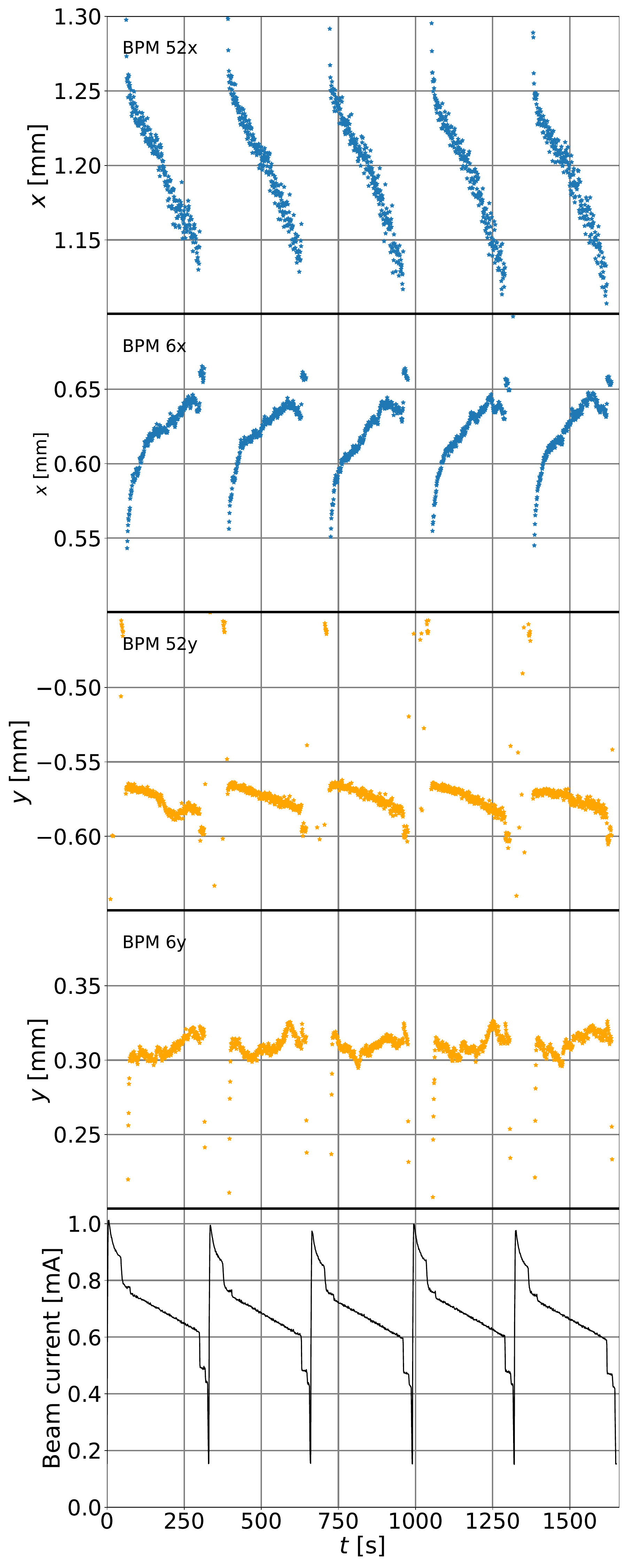}
\caption{\label{fig:COSY-5cycles-xy} Measured horizontal (52x) and
  vertical beam positions (52y) during five consecutive cycles in COSY. For
  comparison the horizontal (vertical) 
  position measurement of the next conventional COSY BPM (6x,6y) $\SI{4}{m}$ downstream in beam
  direction is shown. The lower plots show the beam current. 
  Note that the drifts are in opposite directions which is caused by quadrupole magnets in between the two devices.}
\end{figure}
The Rogowski BPM was used during several experimental beam times by the JEDI
collaboration in order to determine the position of the COSY deuteron beam.
Figures~\ref{fig:COSY-5cycles-xy} show the measured beam position in horizontal
and vertical direction for five consecutive cycles for the Rogowski coil
(52x, 52y) and a conventional COSY BPM (6x, 6y).

\section{Conclusion and outlook}
\label{sec:conclusion}
We described the development of a new type of beam position
monitor based on a Rogowski coil. It reaches a resolution of $\SI{5}{\micro
  m}$ in $\SI{1}{s}$ time interval. The absolute accuracy is determined by
the alignment accuracy of about $\SI{50}{\micro m}$. One advantage over classical
capacitive BPMs is the short insertion length of these devices. 
In an energy-variable machine, 
the frequency-dependent amplification
of the Rogowski BPMs limits its sensitivity, because away from the
resonance the amplitude decreases. In the specific case of a fixed-energy EDM machine, however, this
presents an advantage because, with optimized design, the resonance amplifies
the signals.

While our development uses four toroidal coils, in principle, the device could be divided into more segments in order to study higher-order
components of the transverse particle beam distribution. Note that with the
four-fold segmentation presented here a quadrupole component could already be
studied. A pair of higher
segmented Rogowski coils at a distance in the storage ring may offer the
possibility to measure the phase space distribution of the particle beam.

The individual quarter coils have been measured separately in our investigations to yield the required differential voltage electronically after subtraction. A better read out scheme may be established by a direct galvanic connection between opposing quarters in order to inherently be sensitive to the differential signal without having to deal with the dominant contribution of the beam current represented by the term $D_0(b)$ in Eq.~(\ref{eq:U0-to-lowest-order}). This may also entail the development of a dedicated pre-amplifier for this purpose. Also, the cross talk between neighboring coils due to the induced currents can be reduced. 

\begin{acknowledgments}
This work has been performed in the framework of the JEDI collaboration, and is financially supported by an ERC Advanced-Grant (srEDM \# 694390) of the European Union. 

We are very grateful to J\"urgen B\"oker, Hans-Joachim Krause, Stefan Hintzen, Andrea Pesce, Thomas Sefzick and Jamal Slim
for their support.

\clearpage

\end{acknowledgments}

\begin{appendix}

\section{Detailed derivations}
\label{app:integrals}

Involving Stokes' theorem, the calculation of the magnetic flux due to the beam current enclosed by the winding surfaces $S$ of a quarter coil with index $M = 0, 1, 2, 3$ can be expressed by the line integral of the vector potential over the boundary $\partial S$ of the surface $S$, 
\begin{equation}
    \Phi = \iint_S \vec B \cdot \dd \vec S = \oint_{\partial S} \vec A \cdot \dd \vec \ell \, .
\label{eq:flux-from-vector-potential}
\end{equation}
The approach using the vector potential seems to be advantageous compared to the use of the $B$-field, because in this case, we only have to deal with the single component $A_z$ in order to obtain the enclosed magnetic flux. Furthermore, via Stokes' theorem, one integration less is required. 

\subsection{Derivation of magnetic flux induced in a quadrant coil}

\begin{figure}[h!]
\begin{center}
\includegraphics[width=1\columnwidth]{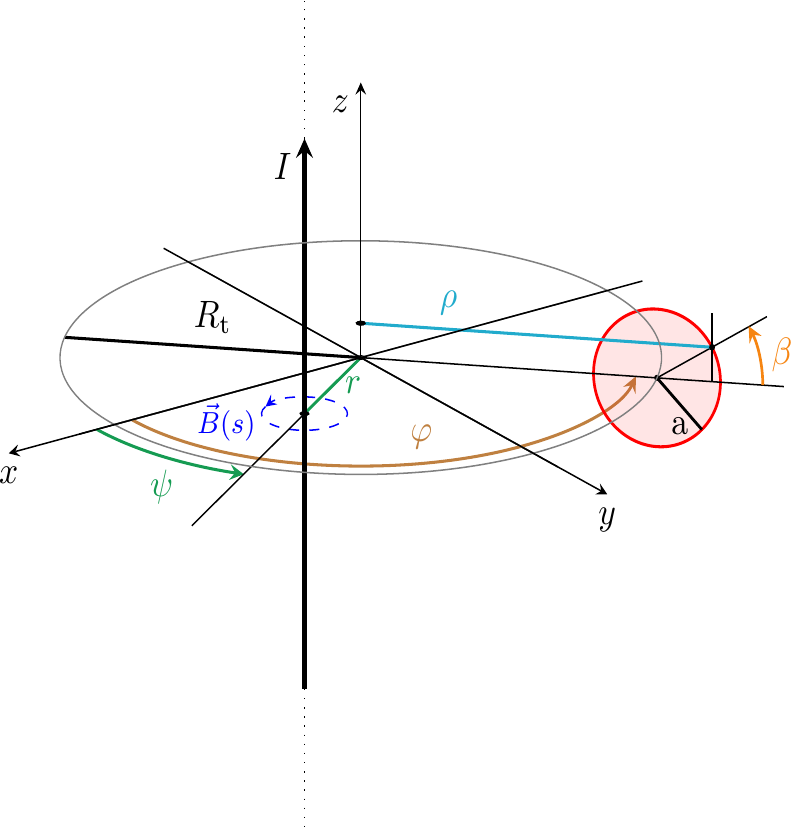}
\caption{\label{fig:position-response-appendix} Representation of the beam coordinates $x$ and $y$, as given in  cylindrical coordinates in  Eq.\,(\ref{eq:parametrization-of-displacements}), and the location of one specific coil winding. The $z$ axis points along the beam direction. The coil center coincides with the center of the coordinate system in the $xy$ plane, the toroid is described by the two radii $R_\textbf{t}$ and $a$. The two variables used for the integration are $\varphi$ and $\beta$. The functional dependence of $\rho(\beta$) is given in Eq.\,(\ref{eq:rho-as-fct-of-beta}).}
\end{center}
\end{figure}

The surface of the torus with given values for $R_\text{t}$ and $a$, as shown in Fig.\,\ref{fig:position-response-appendix},  is parameterized in terms of the two angles $\varphi$ and $\beta$ according to
\begin{equation}
\vec \ell = 
\left(
\begin{array}{c}
\ell_x(\beta, \varphi)\\ 
\ell_y(\beta, \varphi) \\ 
\ell_z(\beta)
\end{array}
\right) = 
\left(
\begin{array}{c}
(R_\text{t} + a \cos\beta) \cdot \cos\varphi\\  
(R_\text{t} + a \cos\beta) \cdot \sin\varphi\\
a \sin\beta
\end{array}
\right)
\,.
\end{equation}
The wire does not cover the surface of the torus entirely but follows a path given by a linear relation between the two angles $\varphi$ and $\beta$
\begin{equation}
\varphi (\beta) = u \cdot \beta \, .
\end{equation}
Due to the fact that the vector potential has only a component along the $z$-direction,
we only need the $z$-component of the differential line element $\dd \vec \ell$ along the coil winding, which is obtained by differentiation
\begin{equation}
 \dd \ell_z (\beta) = a\cos\beta\, \dd \beta\, .
\end{equation}

Furthmore one has to consider that the coils do not cover the full torus. The angles involved are indicated in Fig.~\ref{fig:foto-coil}.

The integral in Eq.\,(\ref{eq:flux-from-vector-potential}) can be further simplified, yielding for the flux induced in a quadrant $M$,
\begin{eqnarray}
\label{eq:master-equation-magnetic-flux}
\lefteqn{\Phi_M(x,y) 
 = \oint \vec A \cdot \dd \vec \ell 
  = \int A_z(\rho(\beta), \varphi(\beta)) \, \dd \ell_z} \nonumber \\
&  =& a \int\displaylimits_{0}^{2\pi \cdot N_\text{w}} A_z(\rho(\beta), \varphi(\beta))  \cos \beta \, \dd \beta \\
  &= &a\frac{2 N_\text{w}}{\pi} \int\displaylimits_{M (\pi/2) + \theta}^{(M+1)  (\pi/2) - \theta} \,\,
\int\displaylimits_0^{2\pi}A_z(\rho(\beta),\varphi)\,\cos \beta \, \dd \beta \, \dd \varphi \, , \nonumber
\end{eqnarray}
where 
\begin{equation}
\rho(\beta) = R_\text{t} + a  \cos\beta = R_{\text{t}} \left(1+ b \cos\beta\right) \,,
\label{eq:rho-as-fct-of-beta}
\end{equation}
as shown in Fig.\,\ref{fig:position-response-appendix}. 
In the last line of Eq.~\ref{eq:master-equation-magnetic-flux}, the sum over the $N_\text{w}$ discrete angle values of $\varphi$ is represented by an integral, which is an adequate description for a densely wound coil. In cylindrical coordinates, the vector potential $A_z(\rho, \varphi)$ has already been introduced in Eq.\,(\ref{eq:Az-of-rho-and-phi}), in which the beam displacement from the center is parameterised in cylindrical coordinates
\begin{equation}
x = r \cos \psi \quad \text{and} \quad y = r \sin \psi\,.
\label{eq:parametrization-of-displacements}
\end{equation}
%

Eq.\,(\ref{eq:Az-of-rho-and-phi}) can be written as
\begin{eqnarray}
\lefteqn{A_z(\rho, \varphi)  = -\frac{\mu_0I}{2\pi}\ln\sqrt{\rho^2 -2\rho r \cos(\varphi - \psi) + r^2 }} \nonumber \\
& =& -  \frac{\mu_0 I}{2\pi} \left[\ln{\rho}  -  \sum_{m=1} ^\infty  \frac{1}{m} \left(\frac{r}{\rho}\right)^m \cos \left[m (\varphi-\psi) \right]   \right] \,,
\label{eq:A_z-of-rho-and-phi}
\end{eqnarray}
where the latter part describes the decomposition into a Fourier series\cite[Eq.\,3.152]{jackson1975classical}, for $r<
\rho$. For a beam current monitor without segmentation of the toroidal coil, the integration over $\varphi$ must be performed from 0 to $2\pi$. Eq.~(\ref{eq:A_z-of-rho-and-phi}) shows that all the cosine terms yield vanishing contributions in this
case. Therefore, the information about the beam displacement encoded in $\psi$ (see Eq.\,(\ref{eq:parametrization-of-displacements}) vanishes as well. A full Rogowski coil without segmentation  therefore constitutes a beam current monitor, its sensitivity being entirely described by the last logarithmic term in Eq.\,(\ref{eq:A_z-of-rho-and-phi}), which depends only on the coil geometry. 

For a toroidal coil segmented into four quadrants $M$, each covering  $\Delta \theta \approx \SI{64}{\degree}$, however, the first integration over $\varphi$ required in Eq.\,(\ref{eq:master-equation-magnetic-flux}) yields
\begin{widetext}
\begin{eqnarray}
\lefteqn{A^\text{(1)}_z(\rho,M) 
 =  \frac{2N_\text{w}}{\pi} \int\displaylimits_{M (\pi/2) + \theta}^{(M+1) (\pi/2) - \theta} A_z(\rho, \varphi)\,  \dd \varphi} \nonumber \\
& =& -\frac{\mu_0 N_\text{w} I}{\pi^2} \left[\Delta\theta \ln (\rho)  - \sum_{m=1}^\infty \frac{1}{m} \left(\frac{r}{\rho}\right)^m 
\left(\int\displaylimits_{M (\pi/2) + \theta}^{(M+1) (\pi/2) - \theta} \cos(m\varphi) \, \dd \varphi \cdot \cos(m \psi) + \int\displaylimits_{M (\pi/2) + \theta}^{(M+1) (\pi/2) - \theta} \sin(m\varphi) \, \dd \varphi  \cdot \sin(m\psi) \right)
\right] \nonumber \\
& =& -\frac{\mu_0 N_\text{w} I}{\pi^2} \left[ \Delta\theta \ln \rho - \sum_{m=1}^\infty \frac{1}{m^2}\left(\frac{r}{\rho}\right)^m \left(c_{m,M} \cos(m\psi) + d_{m,M} \sin(m\psi)   \right)   
\right]\,,
\end{eqnarray}
\end{widetext}
where 
the coefficients 
\begin{eqnarray}
c_{m,M} & = &\sin\left( m \pi\frac{M+1}{2} - \theta \right) - \sin\left(m\pi \frac{M}{2} + \theta  \right)  \label{eq:cs-and-ds}\\
d_{m,M} & = &-\cos\left( m \pi\frac{M+1}{2} - \theta \right) + \cos\left(m\pi \frac{M}{2} + \theta \right)\, . \nonumber
\end{eqnarray}
are functions of the angle $\theta$
 (see Table\,\ref{table:Dm-cs-and-ds}). The dependence on $m$ of the arguments of the trigonometric functions can be eliminated using a sum over the sine and cosine functions, related to the beam displacements $x$ and $y$, via Eq.\,(\ref{eq:parametrization-of-displacements}).  It follows with these trigonometric relations from \cite[1.331-1 and 1.331-3]{Gradshteyn:1702455} that 
\begin{widetext}
\begin{equation}
\begin{split}
A^\text{(1)}_z(\rho,M) = -  \frac{\mu_0 N_\text{w} I}{\pi^2} 
\Biggr[ &\Delta\theta  \ln \rho -  \sum_{m=1}^\infty \frac{1}{m^2} \left( \frac{r}{\rho} \right)^m 
\Biggr(  c_{m,M}  \sum_{n=0}^{\lfloor\frac{m}{2}\rfloor}   \binom{m}{2n}   \left( \cos\psi\right)^{m-2n}       \left( \sin\psi \right)^{2n}   (-1)^n  \\
      + \, & d_{m,M}  \sum_{n=0}^{\lfloor\frac{m-1}{2}\rfloor} \binom{m}{2n+1} \left( \cos\psi\right)^{m-(2n+1)}   \left( \sin\psi \right)^{2n+1} (-1)^n 
       \Biggr) \Biggr]\,,
\end{split}
\end{equation}
\end{widetext}
where, e.g., $\lfloor m \rfloor \equiv \text{floor}(m)$. In terms of the displacements, parameterized using Eq.\,(\ref{eq:parametrization-of-displacements}), one can then write 
\begin{widetext}
\begin{equation}
	\begin{split}
	A^\text{(1)}_z(\rho,M) = - \frac{\mu_0 N_\text{w} I}{\pi^2} 
	\Biggr[ &\Delta\theta  \ln \rho -  \sum_{m=1}^\infty \frac{1}{m^2} \left( \frac{R_t}{\rho} \right)^m 
	\Biggr(  c_{m,M}  \sum_{n=0}^{\lfloor\frac{m}{2}\rfloor}   \binom{m}{2n}   \left( \frac{x}{R_t} \right)^{m-2n}     \left( \frac{y}{R_t} \right)^{2n}   (-1)^n  \\
		  + \, & d_{m,M}  \sum_{n=0}^{\lfloor\frac{m-1}{2}\rfloor} \binom{m}{2n+1} \left( \frac{x}{R_t} \right)^{m-(2n+1)} \left( \frac{y}{R_t} \right)^{2n+1} (-1)^n 
    \Biggr) \Biggr]\,.
\end{split}
\end{equation}
\end{widetext}

The missing integration over $\beta$ according to Eq.\,(\ref{eq:master-equation-magnetic-flux}), taking into account Eq.\,(\ref{eq:rho-as-fct-of-beta}),  yields the magnetic flux
\begin{widetext}
\begin{eqnarray}
\Phi_M(x,y) & = &a\int\displaylimits_{0}^{2\pi} A^\text{(1)}_z(R_\text{t} + a  \cos \beta, M) \cos \beta \, \dd \beta \nonumber \\
&=&  - \frac{\mu_0 N_\text{w} I a}{\pi^2} \left[ \Delta\theta
       A^{(2)} -  \sum_{m=1}^\infty \frac{1}{m^2} \int\displaylimits_{0}^{2\pi}  \left( \frac{R_t}{R_\text{t} + a  \cos\beta} \right)^m \cos \beta \, \dd \beta
       \, \cdot \,E_{m,M}(x,y) \right]\,.
\label{eq:first-integral}
\end{eqnarray}
\end{widetext}
The position-dependent functions $E_{m,M}(x,y)$ are defined below in Eq.~(\ref{eq:Ems-compact}).
The integral in Eq.\,(\ref{eq:first-integral})
can be solved by partial integration and the use of reference~\cite{Gradshteyn:1702455} (3.644-4)
\begin{eqnarray}
A^{(2)} & =& \int\displaylimits_{0}^{2\pi} \ln (R_\text{t} + a  \cos\beta)   \cos \beta \,  \dd \beta \\
   &=&\int_0^{2\pi} \frac{b\sin^2\beta}{1+b\cos\beta} \dd \beta 
         = \frac{2\pi}{b} \left( 1 -\sqrt{1-b^2}  \right)\,, \nonumber
\label{eq:integral-A2}
\end{eqnarray}
with $b = a / R_\text{t}$.  

The induced flux in segment $M$ can be written as
\begin{widetext}
\begin{equation}
\begin{split}
\Phi_M(x,y) = -\mu_0 N_\text{w} I a \Biggr[  \frac{1-\sqrt{1-b^2}}{b} \, \frac{2\Delta \theta}{\pi} - &\sum_{m=1}^{\infty} \frac{C_m(b)}{(m\pi)^2}    
 \Biggr(
 c_{m,M}  \sum_{n=0}^{\lfloor\frac{m}{2}\rfloor}   \binom{m}{2n}   \left( \frac{x}{R_t} \right)^{m-2n}     \left( \frac{y}{R_t} \right)^{2n}   (-1)^n  \\
+ \,  
& d_{m,M}  \sum_{n=0}^{\lfloor\frac{m-1}{2}\rfloor} \binom{m}{2n+1} \left( \frac{x}{R_t} \right)^{m-(2n+1)} \left( \frac{y}{R_t} \right)^{2n+1} (-1)^n
\Biggr)  \Biggr]\,.
\end{split}\label{eq:master-integrated-flux2}
\end{equation}
\end{widetext}

The remaining integral in Eq.\,(\ref{eq:master-integrated-flux2}),
\begin{equation}
C_m(b) =  \int\displaylimits_{0}^{2\pi} \frac{\cos \beta}{\left( 1 + b\cos \beta \right)^m} \, \dd \beta \,,
\label{eq:Cm-of-b}
\end{equation}
is required for $m \ge 1$ and can be evaluated by recursion, as explained in Sec.\,\ref{sec:Cm-of-b}.

With the substitutions 
\begin{eqnarray}
D_0 (b) &=&  \frac{1-\sqrt{1-b^2}}{b} \, \frac{2\Delta \theta}{\pi} \, , \nonumber\\
D_m(b) &=& - \frac{C_m(b)}{(m\pi)^2}\quad  \mbox{for} \quad m=1,2,3,\dots \, ,
\label{eq:D0-and-Dm}
\end{eqnarray}
the induced flux in quadrant $M$ can be written as

\begin{equation}
\Phi_M(x,y) =  -\mu_0 N_\text{w} I a \left[ D_0 (b) + \sum_{m=1}^{\infty} D_m(b) \, E_{m,M}(x,y) \right]\,.
\label{eq:flux-compact}
\end{equation}
The position-dependent terms  $E_{m,M}(x,y)$ are given by
\begin{widetext}
\begin{equation}
E_{m,M}(x,y) = c_{m,M}   \sum_{n=0}^{\lfloor\frac{m}{2}\rfloor}   \binom{m}{2n}   \left( \frac{x}{R_\text{t}} \right)^{m-2n}     \left( \frac{y}{R_\text{t}} \right)^{2n}   (-1)^n  
+ \,   d_{m,M}   \sum_{n=0}^{\lfloor\frac{m-1}{2}\rfloor} \binom{m}{2n+1} \left( \frac{x}{R_\text{t}} \right)^{m-(2n+1)} \left( \frac{y}{R_\text{t}} \right)^{2n+1} (-1)^n \,.
\label{eq:Ems-compact}
\end{equation}
\end{widetext}
Analytic expressions for the functions $D_m(b)$ and the coefficients $c_{m,M}$ and $d_{m,M}$ are listed in Table\,\ref{table:Dm-cs-and-ds}, and for the functions $E_{m,M}(x,y)$ in Table\,\ref{table:Ems}, up to order $m=6$. 

The symmetry relations for the $E_{m,M}(x,y)$ in the different quadrants $M$, given in Table\,\ref{table:Ems}, apply to all orders in $m$ and are passed on to the induced magnetic flux in a quadrant. Thus, given a magnetic flux in quadrant $M=0$ of $\Phi_0 (x,y)$,  one can write for the induced fluxes in the other quadrants,
\begin{eqnarray}
\Phi_1(x,y) & =& \Phi_0(-x,y) \, , \nonumber \\
\Phi_2(x,y) & =& \Phi_0(-x,-y) \, ,  \label{eq:symmetry-relation-for-induced-flux}\\
\Phi_3(x,y) & =& \Phi_0(x,-y) \, \nonumber .
\end{eqnarray}
Please note the equivalence to Eq.~(\ref{eq:symmetry-relation-for-induced-voltages}).

\begin{table*}[tb]
	\caption{\label{table:Dm-cs-and-ds} Analytical expressions for the terms $D_m(b)$ (Eq.~(\ref{eq:D0-and-Dm})) and the coefficients $c_{m,M}$ and $d_{m,M}$, defined in Eq.\,(\ref{eq:cs-and-ds}), for the four quadrants $M$ up to order $m=6$ for a coil covering 
    an angle $\Delta \theta$ starting at $\theta$, i.e. ${\pi}/{2} = \Delta \theta + 2 \theta$. For small values of $b$, $D_m(b) \approx b/(m\pi)$.}
	\begin{ruledtabular}
		\renewcommand{\arraystretch}{2.3}
		\begin{tabular}{cccccccccc}
			$M$  &  &  0 & 1 & 2 & 3 & 0 & 1 & 2 & 3\\\hline
   			$m$  & $D_m(b)$ $ $ & $c_{m,0}$ & $c_{m,1}$ & $c_{m,2}$ & $c_{m,3}$ 
    &  $d_{m,0}$ & $d_{m,1}$ & $d_{m,2}$ & $d_{m,3}$\\ \hline
			1     & $\frac{\displaystyle 2 }{\displaystyle b\pi} \,{\left(\frac{\displaystyle 1}{\displaystyle  \sqrt{1-b^2 }}-1\right)}$& $\cos{\theta}-\sin{\theta}$  &  $-c_{1,0}$ &  $-c_{1,0}$ &  $c_{1,0}$ 
  & $c_{1,0}$ &  $c_{1,0}$ & $-c_{1,0}$ &  $-c_{1,0}$ \\
			2    & $\frac{\displaystyle b}{\displaystyle  {2\pi{\left(1-b^2 \right)}}^{3/2} }$ & $0$  & $0$  & $0$  & $0$ & 
   $2\cos(2\theta) $ &  $-d_{2,0}$ & $d_{2,0}$  & $-d_{2,0}$ \\
			3    & $\frac{\displaystyle b}{\displaystyle  {3\pi{\left(1-b^2 \right)}}^{5/2} }$& $-\cos(3\theta)-\sin(3\theta)$  &$-c_{3,0}$  &    $-c_{3,0}$ &   $c_{3,0}$   &$-c_{3,0}$  &    $-c_{3,0}$ &   $c_{3,0}$ & $c_{3,0}$     \\
			4   &  $\frac{\displaystyle  b\,{\left(4+b^2 \right)}}{\displaystyle  { 16 \pi {\left(1-b^2 \right)}}^{7/2} }$ & $-2 \sin(4\theta)$  &   $c_{4,0}$ &  $c_{4,0}$  &   $c_{4,0}$ & 0 & 0 & 0 & 0 \\
			5   & $\frac{\displaystyle  b\,{\left(4+3\,b^2 \right)}}{\displaystyle 20 \pi {{\left(1-b^2 \right)}}^{9/2} }$  & $\cos(5\theta) - \sin(5\theta)$  & $-c_{5,0}$ &$-c_{5,0}$  &  $c_{5,0}$ &
  $c_{5,0}$ & $c_{5,0}$ & $-c_{5,0}$ & $-c_{5,0}$  \\
			6   &  $\frac{\displaystyle b\,{\left(8+12\,b^2 +b^4\right)}}{\displaystyle 48 \pi \,{{\left(1-b^2 \right)}}^{11/2} }$ & $0$  & $0$  & $0$  & $0$  &
   $2\cos(6\theta)$ & $-d_{6,0}$ &  $d_{6,0}$ & $-d_{6,0}$ \\
  \end{tabular}	
	\end{ruledtabular}

\end{table*}

\begin{table*}[tb]
\caption{\label{table:Ems} Analytical expressions for the terms $E_{m,M}(x,y)$ from Eq.\,(\ref{eq:Ems-compact}) for the four quadrants $M$ up to order $m=6$ for partial coverage of the quadrants. }
\begin{ruledtabular}
	\renewcommand{\arraystretch}{2.3}
	\begin{tabular}{ccccc}
		$M$    & 0    & 1 & 2 & 3                              \\\hline
		   & $E_{m,0}(x,y)$       
	    & $E_{m,1}(x,y)$      
	      & $E_{m,2}(x,y)$       
		   & $E_{m,3}(x,y)$         \\\hline
		1      & $\frac{ \displaystyle x+y}{\displaystyle R_\text{t}}$                               $  \left(\cos(\theta)-\sin(\theta) \right) $   & $E_{1,0}(-x,y)$  & $E_{1,0}(-x,-y)$  & $E_{1,0}(x,-y)$ \\
		2      & $\frac{\displaystyle 4xy}{\displaystyle R_\text{t}^2}$           $ \cos(2\theta) $   & $E_{2,0}(-x,y)$  & $E_{2,0}(-x,-y)$  & $E_{2,0}(x,-y)$  \\
		3      & $\frac{\displaystyle 3\,x\,y^2 -y^3 -x^3 +3\,x^2 \,y}{\displaystyle R_\text{t}^3}$   $(-\cos(3\theta) - \sin(3\theta)) $   & $E_{3,0}(-x,y)$  & $E_{3,0}(-x,-y)$  & $E_{3,0}(x,-y)$ \\
		4        & $ -2 \, \frac{\displaystyle x^4 - 6x^2y^2 +y^4}{\displaystyle R_\text{t}^4}  \,  \sin(4\theta) \,$   & $E_{4,0}(-x,y)$  & $E_{4,0}(-x,-y)$  & $E_{4,0}(x,-y)$            \\
		5      & $\frac{\displaystyle x^5 +y^5 +5\,x\,y^4 +5\,x^4 \,y -10\,x^2 \,y^3 -10\,x^3 \,y^2 }{\displaystyle R_\text{t}^5 }$   $ (\cos(5\theta) - \sin(5\theta)) $  & $E_{5,0}(-x,y)$  & $E_{5,0}(-x,-y)$  & $E_{5,0}(x,-y)$  \\
		6      & $\frac{\displaystyle 12\,x\,y^5 + 12\,x^5 \,y - 40\,x^3 \,y^3 }{\displaystyle R_\text{t}^6 }$     $ \cos(6\theta)   $  & $E_{6,0}(-x,y)$  & $E_{6,0}(-x,-y)$  & $E_{6,0}(x,-y)$ \\ 
	\end{tabular}	
\end{ruledtabular}
 
\end{table*}

\subsection{Recursion formula for $C_m(b)$}
\label{sec:Cm-of-b}

The difference of two consecutive functions is found to be proportional to the derivative of the first one. 
	\begin{eqnarray}
	&C_{m+1}(b) &- C_m(b) \nonumber \\  
	&=& \int\displaylimits_{0}^{2\pi}  \frac{\cos \beta}{(1 + b \cos \beta)^{m}}\left(\frac{1}{1 + b\cos \beta}  -1 \right)  \dd \beta \nonumber \\
	&=& -b \int\displaylimits_{0}^{2\pi} \frac{\cos^2 \beta}{(1 + b \cos \beta)^{m+1}} \, \dd \beta = \frac{b}{m} \frac{\dd}{\dd b} C_m(b)\, . \nonumber \\
 &&
	\end{eqnarray}
Thus, the $(m+1)$-th function can be obtained from the $m$-th function by recursion via 
\begin{equation}
C_{m+1}(b) = C_m(b) + \frac{b}{m}  \frac{\dd}{\dd b} C_m(b)\,,
\end{equation}
where, following Eq.\,(\ref{eq:Cm-of-b}), for $ b \in (-1,1)$ the function starting the recursion is found to be
\begin{equation}
C_1(b)   = \int\displaylimits_{0}^{\pi} \frac{2\cos\beta}{1 + b\cos \beta} \dd \beta 
= \frac{2\pi}{b} \left( 1 - \frac{\displaystyle 1}{\displaystyle \sqrt{1-b^2}}     \right)\,.
\label{eq:C1-of-b}
\end{equation}

\subsection{Magnetic flux in one quadrant without displacement}

Without displacements of the beam with respect to the center of the toroidal coil, \textit{i.e.}, for $x=y=0$, the magnetic flux in Eq.\,(\ref{eq:flux-compact}) is entirely given by the first term, yielding the well-known result
\begin{equation}
\begin{split}
\Phi_M(r=0) & = -\mu_0 N_\text{w} I a \, D_0(b) \\
            & \approx -\mu_0 N_\text{w} I a \frac{b}{2} \\ & = -N_\text{w} \cdot \frac{\mu_0 I}{2 \pi R_\text{t}} \cdot \pi a^2 = -N_\text{w} \cdot B \cdot S
            \,,
\end{split}
\label{eq:magnetic-flux-without-displacement}
\end{equation}
that the magnetic flux is equal to the product of the number of coil windings $N_\text{w}$, magnetic flux density $B$ at a distance of $R_\text{t}$ and the enclosed winding area $S$. (In the second line, we have made use of the fact that $b \ll 1$, see Table\,\ref{table:basic-parameters}.)

\section{Resonance frequency of quarter toroid}
\label{app:resonance-frequency}

The numerical calculation of the resonance frequency is carried out using the equivalent circuit diagram shown in Fig.\,\ref{fig:equivalent-circuit}. 
%
The relation of the measured output voltage $U_M$ to the beam induced voltage $U_M^{\text{ind}}$ can be described in the frequency domain using the picture of a voltage divider, which leads to
\begin{equation}
\begin{split}
\frac{U_M}{U^{\text{ind}}_M} & = \frac{\left( \frac{1}{R_\text{out}} + i\omega C\right)^{-1}}  {i \omega L +R + \left( \frac{1}{R_\text{out}} + i\omega C\right)^{-1}} \\
& = \frac{1}{(i \omega L + R) \left( \frac{1}{R_\text{out}} + i \omega C \right) +1}\,.
\end{split}
\end{equation}
The amplitude can be written as
\begin{eqnarray}
 \lefteqn{F(\omega)  = \left| \frac{U_M}{U^{\text{ind}}_M} \right|}\nonumber \\
& = &\frac{1}{\sqrt{\left (1 - \omega^2 L C + \frac{R}{R_\text{out}} \right)^2 + \left(\frac{\omega L}{R_\text{out}} + \omega R C \right)^2}}\,,
\label{eq:f-of-omega}
\end{eqnarray}
and the phase is given by:
\begin{equation}
\phi(\omega) = \arctan\left( 
\frac{\omega ( C R R_{\text{out}} + L)}{C L R_{\text{out}} \omega^2 - R - R_{\text{out}}}
\right)    \, .
\label{eq:phi-of-omega}
\end{equation}

To estimate $C$ and $L$ in this circuit we consider 
the inner and outer radii of the toroidal coil
\begin{equation}
\begin{split}
    r_\text{i} & = R_\text{t} - a = \SI{52.5}{mm} \, ,\\
    r_\text{o} & = R_\text{t} + a = \SI{64.5}{mm}\,, 
 \end{split}
\end{equation}
using $R_\text{t}$ and $a$, as listed in Table\,\ref{table:basic-parameters}. 

The nominal outer wire diameter $d_\text{w}$ is given by
\begin{equation}
d_\text{w} = 2r_\text{w} = \SI{450}{\micro \meter}\,,
\label{eq:wire-diameter}
\end{equation} 
its insulation is $\SI{50}{\micro \meter}$ thick, i.e., the diameter of the current-carrying wire is $\SI{400}{\micro \meter}$. The angle between two wire centers seen from the center of the toroid amounts to
\begin{equation}
 \alpha = 2  \arcsin \left( \frac{\textstyle  r_\text{w}}{\textstyle r_\text{i} - r_\text{w}}\right)\,.
\end{equation}
The maximum number of windings of a fully wound quadrant coil is therefore given by
\begin{equation}
 N_\text{w}^\text{fq} = \text{floor} \left(\frac{\pi}{2\alpha} \right) = 182\,.
 \label{eq:number-of-windings}
\end{equation}

In order to estimate the inductance of a quarter coil, we quote the textbook formula for a long, straight and densely wound coil
\begin{equation}
L^\text{fq} 
= \mu_0 \mu_r \frac{\left(N^\text{fq}_\text{w}\right)^2}{\ell} S 
\approx \SI{57.09}{\micro\henry}\,,
\label{eq:inductance-of-coil}
\end{equation}
where $N^\text{fq}_\text{w}$ is the number of windings of the quarter toroid, and $\ell = \frac{\pi}{2}r_\text{i}$  is  the length of the coil along its inner radius. $S$ denotes the cross section of the coil. Here the permeability of vacuum is
\begin{equation}
\mu_0  = 1.25663706127(20) \cdot  \SI{e-6} {\volt \second\per \ampere \per \meter}  
\end{equation}
and the relative permeability $\mu_r$ of the material of the coil body  is close to unity.

As shown in Fig.\,\ref{fig:foto-coil}, the coils do not cover the quadrants completely, and instead of $N_\text{w}$,  given in Eq.\,(\ref{eq:number-of-windings}), the real number of windings used in a quarter coil amounted to 
\begin{equation}
N_\text{w} = 132\,.
\end{equation}
The corresponding real inductance of a partially wound coil with the above given number of windings corresponds to
\begin{equation}
L =  \frac{N_\text{w}}{N^\text{fq}_\text{w}} \cdot 
 L^\text{fq} \approx \SI{41.40}{\micro\henry}\,.
\end{equation}

The capacitances of the twisted connecting wires are estimated like in a parallel plate capacitor using a length
and a plate separation of 
\begin{equation}
  \ell_C = \SI{5}{cm} \quad \text{and} \quad d_C = 0.1\cdot d_\text{w} \,,
\end{equation}
yielding
\begin{equation}
 C_{tw} = \epsilon_0 \cdot \frac{\ell_C \cdot d_\text{w} }{d_C} \approx \SI{4.4}{\pico \farad}\,,
\end{equation}
where the vacuum permittivity is
\begin{equation}
 \epsilon_0  \approx  \SI{8.854e-12}{\ampere \second \per \volt \per \meter}\,.
\end{equation}
These twisted wires are soldered to a \SI{17}{cm} long coaxial cable with an specific capacity of \SI{96}{pF/m}, resulting
in a capacity of $C_{cc}=\SI{16.3}{pF}$.
This ammounts to a total capcity
\begin{equation}
    C = C_{tw}+C_{cc} = \SI{20.7}{pF} \, .
\end{equation}

An estimate of the resonance frequency of a quarter of a Rogowski coil using the specified geometrical and physical parameters given above, yields 
\begin{equation}
   f_0 = \frac{1}{2 \pi \sqrt{L C} } \approx \SI{5.44}{\mega \hertz}\,.
 \label{eq:f0-numerical-value}
\end{equation}
The resistance of a coil quarter amounts to 
\begin{equation}
 R = \frac{N_\text{w} \cdot 8(a+d_\text{w})}{\sigma_\text{Cu} \cdot  {d_\text{w}}^2} \approx \SI{0.61}{\ohm}\,,
\end{equation}
where for the conductivity of copper after winding, a value of
\begin{equation}
 \sigma_\text{Cu} = \SI{55e6}{\siemens \per \meter}
\end{equation}
was used.

\begin{figure}[t!]
	\begin{center}
  \includegraphics[width=\columnwidth]{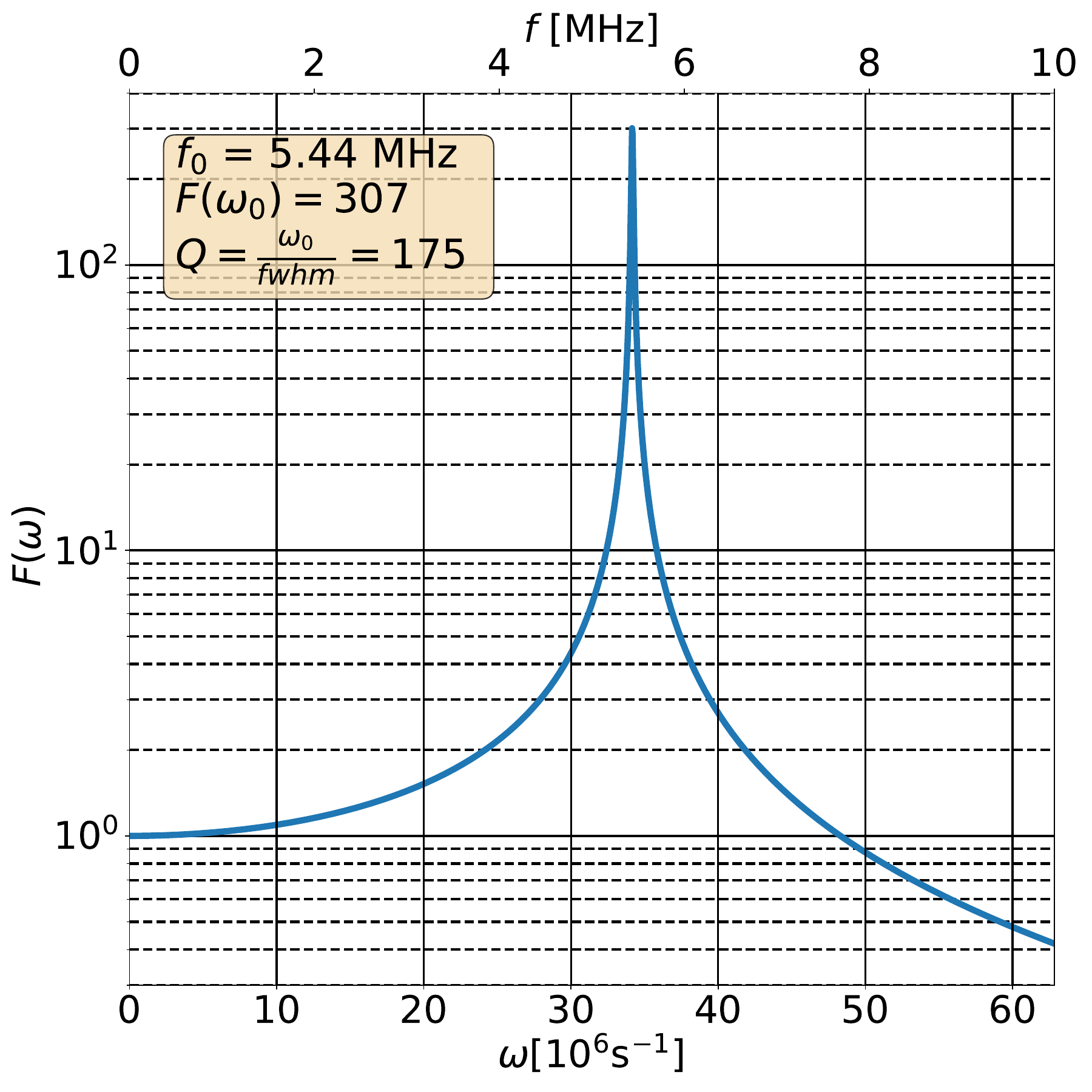}
	\end{center}
	\begin{center}
		\caption{\label{fig:resonance-frequency} Plot of the frequency response function $F(\omega)$ from  Eq.\,(\ref{eq:f-of-omega}) using $R_\text{out}$, $R$, $C$, and $L$ given in Table\,\ref{table:basic-parameters}. The resonance frequency $f_0$ is given in Eq.\,(\ref{eq:f0-numerical-value}), and the estimated quality factor $Q \approx 175$.}
	\end{center}
\end{figure}

The function $F(\omega)$ from Eq.\,(\ref{eq:f-of-omega}) is plotted in Fig.\,\ref{fig:resonance-frequency} using the values for $C$, $L$, $R$, and $R_\text{out}$ listed in Table\,\ref{table:basic-parameters}. For an input impedance of $R_\text{out} = \SI{500}{\kilo \ohm}$, a quality factor for the operation of a quarter coil near the resonance $\omega_0$ of
\begin{equation}
Q = \frac{\omega_0}{\text{fwhm}} \approx 175\,
\label{eq:Q-value-resonance}
\end{equation}
can be estimated.

\vspace*{5mm}

\section{Effect of mirror currents}\label{app:mirror_currents}
The distribution of the image current density for a pencil beam depends on its position ($r$, $\psi$) \cite{wendt2020bpm}.
\begin{eqnarray}
   j (\rho,\phi) &=& \frac{-I}{2\pi \rho} \, \frac{\left(1-\alpha^2\right)\delta(\rho-u)}{1-2\alpha\cos(\phi-\psi)+\alpha^2} \\
   &=& \frac{-I}{2\pi\rho}\left[1+2\sum_{m=1}^\infty \alpha^m \cos(m(\phi-\psi)) \right] \delta(\rho-u) \, , \nonumber \\
   &&\mbox{with} \quad \alpha = \frac{r}{\rho} \, ,\nonumber 
\end{eqnarray}
where the delta-function has been introduced to state that the current density is confined to the inner tube surface because of the skin effect.  The inner radius of the beam tube is denoted by $u$, and the beam coordinates $r$ and $\psi$ are defined in Eq.~(\ref{eq:parametrization-of-displacements}). On integrating one finds that the total current on the wall matches exactly the beam current $I$.

A pick-up coil like a segmented Rogowski coil does not only capture the magnetic field of the beam, but also that of the wall currents, which constructively adds to the primary contribution by the beam, because they flow on the other side of the coils in reverse direction.
The total flux through a coil due to the wall currents can be found similar to the discussion of Eq.~(\ref{eq:flux-from-vector-potential}). Equation~(\ref{eq:Az-of-rho-and-phi}) is a special case of the more general expression of the vector potential for a two-dimensional current distribution
\begin{equation}
A_z(\rho,\phi) = \frac{\mu_0}{2\pi} \int_{R^2} j(\vec{r}\,') \ln(|\vec{r}(\rho,\phi)-\vec{r}\,'|) \dd \vec{r}\,' \, ,
\end{equation}
which for the wall currents takes the form
\begin{widetext}
\begin{eqnarray}
A_z(\rho,\phi) &=& \frac{\mu_0 I}{4\pi^2} \int_0^{2\pi}\int_0^{\infty}  \frac{1}{\rho'} \,
\frac{\left(1-\alpha'^2 \right) \delta(\rho'-s)}
{1-2\alpha'^2 \cos(\phi'-\psi)+\alpha'^2 }
\ln \left(\sqrt{\rho'^2-2\rho'\rho\cos(\phi'-\phi)+\rho^2}\right)\rho'\dd\rho'\dd\phi' \, \\
&&\mbox{with} \quad \alpha' = \frac{r}{\rho'} \, .\nonumber 
\end{eqnarray}
\end{widetext}
In order to find the lowest order term ($m = 1$) in the displacement we can expand both the fraction and the logarithm (see Eq.~(\ref{eq:A_z-of-rho-and-phi}))
\begin{eqnarray}
A_z(\rho,\phi) &=& \frac{\mu_0 I}{8\pi^2} \int_0^{2\pi} \left( 1+2 \frac{r}{u} \cos(\phi'-\psi) \right) \nonumber \\
& & \cdot \left( \ln(u)-\frac{\rho}{u}\cos(\phi'-\phi)\right) \dd\phi'
\end{eqnarray}
and see that only the product of the two cos-terms yields the desired contribution.
\begin{eqnarray}
A_z(\rho,\phi) &=& -\frac{\mu_0 I}{4\pi^2} \frac{r \rho}{u^2} \int_0^{2\pi} \cos(\phi'-\psi) \cos(\phi'-\phi) \dd\phi' \nonumber \\
&= &-\frac{\mu_0 I}{4\pi} \frac{r \rho}{u^2} \cos(\phi-\psi) \, .
\end{eqnarray}
Analogous to Eq.~(\ref{eq:master-equation-magnetic-flux}) the flux induced in the quadrant coil $M$ is obtained by the integration over its surface
\begin{eqnarray}
\Phi_M &=& \frac{2aN_\text{w}}{\pi} \int_{M (\pi/2) - \theta}^{(M+1) (\pi/2) + \theta} \int_0^{2\pi}  \!\!  A_z(\rho(\beta),\beta)\cos(\beta) \dd\beta \dd\phi \nonumber \\
&=& \frac{a\mu_0 N_\text{w} I}{2\pi^2} \frac{r}{u^2} \int_0^{2\pi} [R_\text{t} + a\cos(\beta)] \cos(\beta) \dd \beta  \nonumber \\
& &\cdot \int_{M (\pi/2) + \theta}^{(M+1) (\pi/2) - \theta} \cos(\phi-\psi) \dd \phi \, ,
\end{eqnarray}

\begin{eqnarray}
\Phi_M(x,y) &=& \frac{a\mu_0 N_\text{w} I}{2\pi^2} \frac{r}{u^2}  a \pi [c_{1,M} \cos(\psi) + d_{1,M} \sin(\psi)] \nonumber \\
&=& \frac{\mu_0 N_\text{w} I}{2 \pi} \frac{a^2}{u^2}  [c_{1,M} \, x + d_{1,M} \, y]
\end{eqnarray}
with the coefficients $c_{1,M}$ and $d_{1,M}$ defined in Eq.~(\ref{eq:cs-and-ds}). In particular
\begin{eqnarray}
\Phi_0(x,y) &=& \frac{\mu_0 N_\text{w} I}{2\pi} \frac{a^2}{u^2}(\cos\theta - \sin\theta) (x+y) \nonumber \\
&=& \mu_0 N_\text{w} I a \left( \frac{a R_\text{t}}{2\pi u^2} \right)
E_{1,0}(x,y) \, . \label{eq:phi0_mc}
\end{eqnarray}
This flux adds to the primary one from the beam current, see Eq.~\ref{eq:flux-compact}. There the term in brackets has its beam current counterpart in
\begin{equation}
 D_1(b) = \frac{2}{b\pi} \left(\frac{1}{\sqrt{1-b^2}} -1 \right) \approx \frac{b}{\pi}
\end{equation}
which is about half the value in brackets for a Rogowski quarter coil close to the beam tube  $u \approx R_\text{t}$. This explains in part the about 50\% higher value of the linear term of the induced voltage with respect to the displacement found experimentally, see the discussion following Eq.\ref{c1'-result}.

\end{appendix}
\bibliographystyle{apsrev4-1}
\bibliography{dBase_24.11.2020}

\begin{thebibliography}{26}%
\makeatletter
\providecommand \@ifxundefined [1]{%
 \@ifx{#1\undefined}
}%
\providecommand \@ifnum [1]{%
 \ifnum #1\expandafter \@firstoftwo
 \else \expandafter \@secondoftwo
 \fi
}%
\providecommand \@ifx [1]{%
 \ifx #1\expandafter \@firstoftwo
 \else \expandafter \@secondoftwo
 \fi
}%
\providecommand \natexlab [1]{#1}%
\providecommand \enquote  [1]{``#1''}%
\providecommand \bibnamefont  [1]{#1}%
\providecommand \bibfnamefont [1]{#1}%
\providecommand \citenamefont [1]{#1}%
\providecommand \href@noop [0]{\@secondoftwo}%
\providecommand \href [0]{\begingroup \@sanitize@url \@href}%
\providecommand \@href[1]{\@@startlink{#1}\@@href}%
\providecommand \@@href[1]{\endgroup#1\@@endlink}%
\providecommand \@sanitize@url [0]{\catcode `\\12\catcode `\$12\catcode
  `\&12\catcode `\#12\catcode `\^12\catcode `\_12\catcode `\%12\relax}%
\providecommand \@@startlink[1]{}%
\providecommand \@@endlink[0]{}%
\providecommand \url  [0]{\begingroup\@sanitize@url \@url }%
\providecommand \@url [1]{\endgroup\@href {#1}{\urlprefix }}%
\providecommand \urlprefix  [0]{URL }%
\providecommand \Eprint [0]{\href }%
\providecommand \doibase [0]{http://dx.doi.org/}%
\providecommand \selectlanguage [0]{\@gobble}%
\providecommand \bibinfo  [0]{\@secondoftwo}%
\providecommand \bibfield  [0]{\@secondoftwo}%
\providecommand \translation [1]{[#1]}%
\providecommand \BibitemOpen [0]{}%
\providecommand \bibitemStop [0]{}%
\providecommand \bibitemNoStop [0]{.\EOS\space}%
\providecommand \EOS [0]{\spacefactor3000\relax}%
\providecommand \BibitemShut  [1]{\csname bibitem#1\endcsname}%
\let\auto@bib@innerbib\@empty
\bibitem [{\citenamefont {Pospelov}\ and\ \citenamefont
  {Ritz}(2005)}]{POSPELOV2005119}%
  \BibitemOpen
  \bibfield  {author} {\bibinfo {author} {\bibfnamefont {M.}~\bibnamefont
  {Pospelov}}\ and\ \bibinfo {author} {\bibfnamefont {A.}~\bibnamefont
  {Ritz}},\ }\href {\doibase https://doi.org/10.1016/j.aop.2005.04.002}
  {\bibfield  {journal} {\bibinfo  {journal} {Annals of Physics}\ }\textbf
  {\bibinfo {volume} {318}},\ \bibinfo {pages} {119 } (\bibinfo {year}
  {2005})},\ \bibinfo {note} {special Issue}\BibitemShut {NoStop}%
\bibitem [{\citenamefont {Robson}(2018)}]{robson:2018}%
  \BibitemOpen
  \bibfield  {author} {\bibinfo {author} {\bibfnamefont {B.}~\bibnamefont
  {Robson}},\ }\href {\doibase 10.4236/jhepgc.2018.41015} {\bibfield  {journal}
  {\bibinfo  {journal} {Journal of High Energy Physics, Gravitation and
  Cosmology}\ }\textbf {\bibinfo {volume} {04}},\ \bibinfo {pages} {166}
  (\bibinfo {year} {2018})}\BibitemShut {NoStop}%
\bibitem [{\citenamefont {Bernreuther}(2002)}]{Bernreuther2002}%
  \BibitemOpen
  \bibfield  {author} {\bibinfo {author} {\bibfnamefont {W.}~\bibnamefont
  {Bernreuther}},\ }\enquote {\bibinfo {title} {Cp violation and
  baryogenesis},}\ in\ \href {\doibase 10.1007/3-540-47895-7_7} {\emph
  {\bibinfo {booktitle} {CP Violation in Particle, Nuclear and
  Astrophysics}}},\ \bibinfo {editor} {edited by\ \bibinfo {editor}
  {\bibfnamefont {M.}~\bibnamefont {Beyer}}}\ (\bibinfo  {publisher} {Springer
  Berlin Heidelberg},\ \bibinfo {address} {Berlin, Heidelberg},\ \bibinfo
  {year} {2002})\ pp.\ \bibinfo {pages} {237--293}\BibitemShut {NoStop}%
\bibitem [{\citenamefont {Abusaif}\ \emph {et~al.}(2021)\citenamefont
  {Abusaif}, \citenamefont {Keshelashvili}, \citenamefont {Grigoryev},
  \citenamefont {Mchedlishvili}, \citenamefont {Jorat}, \citenamefont {Pretz},
  \citenamefont {Mei{\ss}ner}, \citenamefont {Kulikov}, \citenamefont {Stahl},
  \citenamefont {Felden}, \citenamefont {Martin}, \citenamefont {Stassen},
  \citenamefont {W{\"u}stner}, \citenamefont {Aksentev}, \citenamefont
  {Javakhishvili}, \citenamefont {Valetov}, \citenamefont {Soltner},
  \citenamefont {{Alberdi-Esuain}}, \citenamefont {Talman}, \citenamefont
  {Shmakova}, \citenamefont {Kacharava}, \citenamefont {H{\"o}lscher},
  \citenamefont {Ciullo}, \citenamefont {M{\"u}ller}, \citenamefont {Silenko},
  \citenamefont {Gebel}, \citenamefont {Lorentz}, \citenamefont {Natour},
  \citenamefont {Grzonka}, \citenamefont {Zurek.}, \citenamefont {Hetzel},
  \citenamefont {Zupranski}, \citenamefont {Siddique}, \citenamefont {Magiera},
  \citenamefont {Ciepa{\l}}, \citenamefont {B{\"o}hme}, \citenamefont
  {Nikolaev}, \citenamefont {Dymov}, \citenamefont {Lehrach}, \citenamefont
  {Gaisser}, \citenamefont {Wro{\'n}ska}, \citenamefont {Heberling},
  \citenamefont {Contalbrigo}, \citenamefont {Kamerdzhiev}, \citenamefont
  {Stephenson}, \citenamefont {Nass}, \citenamefont {Weidemann}, \citenamefont
  {Senichev}, \citenamefont {Schott}, \citenamefont {Ritman}, \citenamefont
  {Wirzba}, \citenamefont {Bey{\ss}}, \citenamefont {Haj~Tahar}, \citenamefont
  {Str{\"o}her}, \citenamefont {Koop}, \citenamefont {Carli}, \citenamefont
  {Lamont}, \citenamefont {Berz}, \citenamefont {B{\"o}ker}, \citenamefont
  {Pesce}, \citenamefont {Tagliente}, \citenamefont {K{\"a}seberg},
  \citenamefont {Saleev}, \citenamefont {Wagner}, \citenamefont {Makino},
  \citenamefont {Aggarwal}, \citenamefont {Slim}, \citenamefont {Prasuhn},
  \citenamefont {Macharashvili}, \citenamefont {Borburgh}, \citenamefont
  {Lomidze}, \citenamefont {Poncza}, \citenamefont {Shergelashvili},
  \citenamefont {Gagoshidze}, \citenamefont {Tabidze}, \citenamefont {Karanth},
  \citenamefont {Laihem}, \citenamefont {Hahnraths}, \citenamefont {Lenisa},
  \citenamefont {Hejny}, \citenamefont {Giese}, \citenamefont {Sefzick},
  \citenamefont {Barion}, \citenamefont {Michaud}, \citenamefont {Nogga},
  \citenamefont {Straatmann}, \citenamefont {Uzikov}, \citenamefont {Basile},
  \citenamefont {Rathmann}, \citenamefont {Atanasov}, \citenamefont
  {Metreveli}, \citenamefont {Rosenthal}, \citenamefont {Valdau}, \citenamefont
  {Simon},\ and\ \citenamefont {De~Conto}}]{cpEDM2021}%
  \BibitemOpen
  \bibfield  {author} {\bibinfo {author} {\bibfnamefont {F.}~\bibnamefont
  {Abusaif}}, \bibinfo {author} {\bibfnamefont {I.}~\bibnamefont
  {Keshelashvili}}, \bibinfo {author} {\bibfnamefont {K.}~\bibnamefont
  {Grigoryev}}, \bibinfo {author} {\bibfnamefont {D.}~\bibnamefont
  {Mchedlishvili}}, \bibinfo {author} {\bibfnamefont {L.}~\bibnamefont
  {Jorat}}, \bibinfo {author} {\bibfnamefont {J.}~\bibnamefont {Pretz}},
  \bibinfo {author} {\bibfnamefont {U.-G.}\ \bibnamefont {Mei{\ss}ner}},
  \bibinfo {author} {\bibfnamefont {A.}~\bibnamefont {Kulikov}}, \bibinfo
  {author} {\bibfnamefont {A.}~\bibnamefont {Stahl}}, \bibinfo {author}
  {\bibfnamefont {O.}~\bibnamefont {Felden}}, \bibinfo {author} {\bibfnamefont
  {S.}~\bibnamefont {Martin}}, \bibinfo {author} {\bibfnamefont
  {R.}~\bibnamefont {Stassen}}, \bibinfo {author} {\bibfnamefont
  {P.}~\bibnamefont {W{\"u}stner}}, \bibinfo {author} {\bibfnamefont
  {A.}~\bibnamefont {Aksentev}}, \bibinfo {author} {\bibfnamefont
  {O.}~\bibnamefont {Javakhishvili}}, \bibinfo {author} {\bibfnamefont
  {E.}~\bibnamefont {Valetov}}, \bibinfo {author} {\bibfnamefont
  {H.}~\bibnamefont {Soltner}}, \bibinfo {author} {\bibfnamefont
  {B.}~\bibnamefont {{Alberdi-Esuain}}}, \bibinfo {author} {\bibfnamefont
  {R.}~\bibnamefont {Talman}}, \bibinfo {author} {\bibfnamefont
  {V.}~\bibnamefont {Shmakova}}, \bibinfo {author} {\bibfnamefont
  {A.}~\bibnamefont {Kacharava}}, \bibinfo {author} {\bibfnamefont
  {D.}~\bibnamefont {H{\"o}lscher}}, \bibinfo {author} {\bibfnamefont
  {G.}~\bibnamefont {Ciullo}}, \bibinfo {author} {\bibfnamefont
  {F.}~\bibnamefont {M{\"u}ller}}, \bibinfo {author} {\bibfnamefont
  {A.}~\bibnamefont {Silenko}}, \bibinfo {author} {\bibfnamefont
  {R.}~\bibnamefont {Gebel}}, \bibinfo {author} {\bibfnamefont
  {B.}~\bibnamefont {Lorentz}}, \bibinfo {author} {\bibfnamefont
  {G.}~\bibnamefont {Natour}}, \bibinfo {author} {\bibfnamefont
  {D.}~\bibnamefont {Grzonka}}, \bibinfo {author} {\bibfnamefont
  {M.}~\bibnamefont {Zurek.}}, \bibinfo {author} {\bibfnamefont
  {J.}~\bibnamefont {Hetzel}}, \bibinfo {author} {\bibfnamefont
  {P.}~\bibnamefont {Zupranski}}, \bibinfo {author} {\bibfnamefont
  {S.}~\bibnamefont {Siddique}}, \bibinfo {author} {\bibfnamefont
  {A.}~\bibnamefont {Magiera}}, \bibinfo {author} {\bibfnamefont
  {I.}~\bibnamefont {Ciepa{\l}}}, \bibinfo {author} {\bibfnamefont
  {C.}~\bibnamefont {B{\"o}hme}}, \bibinfo {author} {\bibfnamefont
  {N.}~\bibnamefont {Nikolaev}}, \bibinfo {author} {\bibfnamefont
  {S.}~\bibnamefont {Dymov}}, \bibinfo {author} {\bibfnamefont
  {A.}~\bibnamefont {Lehrach}}, \bibinfo {author} {\bibfnamefont
  {M.}~\bibnamefont {Gaisser}}, \bibinfo {author} {\bibfnamefont
  {A.}~\bibnamefont {Wro{\'n}ska}}, \bibinfo {author} {\bibfnamefont
  {D.}~\bibnamefont {Heberling}}, \bibinfo {author} {\bibfnamefont
  {M.}~\bibnamefont {Contalbrigo}}, \bibinfo {author} {\bibfnamefont
  {V.}~\bibnamefont {Kamerdzhiev}}, \bibinfo {author} {\bibfnamefont
  {E.}~\bibnamefont {Stephenson}}, \bibinfo {author} {\bibfnamefont
  {A.}~\bibnamefont {Nass}}, \bibinfo {author} {\bibfnamefont {C.}~\bibnamefont
  {Weidemann}}, \bibinfo {author} {\bibfnamefont {Y.}~\bibnamefont {Senichev}},
  \bibinfo {author} {\bibfnamefont {M.}~\bibnamefont {Schott}}, \bibinfo
  {author} {\bibfnamefont {J.}~\bibnamefont {Ritman}}, \bibinfo {author}
  {\bibfnamefont {A.}~\bibnamefont {Wirzba}}, \bibinfo {author} {\bibfnamefont
  {M.}~\bibnamefont {Bey{\ss}}}, \bibinfo {author} {\bibfnamefont
  {M.}~\bibnamefont {Haj~Tahar}}, \bibinfo {author} {\bibfnamefont
  {H.}~\bibnamefont {Str{\"o}her}}, \bibinfo {author} {\bibfnamefont
  {I.}~\bibnamefont {Koop}}, \bibinfo {author} {\bibfnamefont {C.}~\bibnamefont
  {Carli}}, \bibinfo {author} {\bibfnamefont {M.}~\bibnamefont {Lamont}},
  \bibinfo {author} {\bibfnamefont {M.}~\bibnamefont {Berz}}, \bibinfo {author}
  {\bibfnamefont {J.}~\bibnamefont {B{\"o}ker}}, \bibinfo {author}
  {\bibfnamefont {A.}~\bibnamefont {Pesce}}, \bibinfo {author} {\bibfnamefont
  {G.}~\bibnamefont {Tagliente}}, \bibinfo {author} {\bibfnamefont
  {C.}~\bibnamefont {K{\"a}seberg}}, \bibinfo {author} {\bibfnamefont
  {A.}~\bibnamefont {Saleev}}, \bibinfo {author} {\bibfnamefont
  {T.}~\bibnamefont {Wagner}}, \bibinfo {author} {\bibfnamefont
  {K.}~\bibnamefont {Makino}}, \bibinfo {author} {\bibfnamefont
  {A.}~\bibnamefont {Aggarwal}}, \bibinfo {author} {\bibfnamefont
  {J.}~\bibnamefont {Slim}}, \bibinfo {author} {\bibfnamefont {D.}~\bibnamefont
  {Prasuhn}}, \bibinfo {author} {\bibfnamefont {G.}~\bibnamefont
  {Macharashvili}}, \bibinfo {author} {\bibfnamefont {J.}~\bibnamefont
  {Borburgh}}, \bibinfo {author} {\bibfnamefont {N.}~\bibnamefont {Lomidze}},
  \bibinfo {author} {\bibfnamefont {V.}~\bibnamefont {Poncza}}, \bibinfo
  {author} {\bibfnamefont {D.}~\bibnamefont {Shergelashvili}}, \bibinfo
  {author} {\bibfnamefont {M.}~\bibnamefont {Gagoshidze}}, \bibinfo {author}
  {\bibfnamefont {M.}~\bibnamefont {Tabidze}}, \bibinfo {author} {\bibfnamefont
  {S.}~\bibnamefont {Karanth}}, \bibinfo {author} {\bibfnamefont
  {K.}~\bibnamefont {Laihem}}, \bibinfo {author} {\bibfnamefont
  {T.}~\bibnamefont {Hahnraths}}, \bibinfo {author} {\bibfnamefont
  {P.}~\bibnamefont {Lenisa}}, \bibinfo {author} {\bibfnamefont
  {V.}~\bibnamefont {Hejny}}, \bibinfo {author} {\bibfnamefont
  {N.}~\bibnamefont {Giese}}, \bibinfo {author} {\bibfnamefont
  {T.}~\bibnamefont {Sefzick}}, \bibinfo {author} {\bibfnamefont
  {L.}~\bibnamefont {Barion}}, \bibinfo {author} {\bibfnamefont
  {J.}~\bibnamefont {Michaud}}, \bibinfo {author} {\bibfnamefont
  {A.}~\bibnamefont {Nogga}}, \bibinfo {author} {\bibfnamefont
  {H.}~\bibnamefont {Straatmann}}, \bibinfo {author} {\bibfnamefont
  {Y.}~\bibnamefont {Uzikov}}, \bibinfo {author} {\bibfnamefont
  {S.}~\bibnamefont {Basile}}, \bibinfo {author} {\bibfnamefont
  {F.}~\bibnamefont {Rathmann}}, \bibinfo {author} {\bibfnamefont
  {A.}~\bibnamefont {Atanasov}}, \bibinfo {author} {\bibfnamefont
  {Z.}~\bibnamefont {Metreveli}}, \bibinfo {author} {\bibfnamefont
  {M.}~\bibnamefont {Rosenthal}}, \bibinfo {author} {\bibfnamefont
  {Y.}~\bibnamefont {Valdau}}, \bibinfo {author} {\bibfnamefont
  {M.}~\bibnamefont {Simon}}, \ and\ \bibinfo {author} {\bibfnamefont {J.-M.}\
  \bibnamefont {De~Conto}},\ }\href {\doibase 10.23731/CYRM-2021-003}
  {\bibfield  {journal} {\bibinfo  {journal} {CERN Yellow Report}\ ,\ \bibinfo
  {pages} {257}} (\bibinfo {year} {2021})}\BibitemShut {NoStop}%
\bibitem [{\citenamefont {Rathmann}\ \emph {et~al.}(2013)\citenamefont
  {Rathmann}, \citenamefont {Saleev},\ and\ \citenamefont
  {Nikolaev}}]{Rathmann:2013rqa}%
  \BibitemOpen
  \bibfield  {author} {\bibinfo {author} {\bibfnamefont {F.}~\bibnamefont
  {Rathmann}}, \bibinfo {author} {\bibfnamefont {A.}~\bibnamefont {Saleev}}, \
  and\ \bibinfo {author} {\bibfnamefont {N.~N.}\ \bibnamefont {Nikolaev}}
  (\bibinfo {collaboration} {JEDI, srEDM}),\ }\href
  {https://www.doi.org/10.1088/1742-6596/447/1/012011} {\bibfield  {journal}
  {\bibinfo  {journal} {J.\ Phys.\ Conf.\ Ser.}\ }\textbf {\bibinfo {volume}
  {447}},\ \bibinfo {pages} {012011} (\bibinfo {year} {2013})}\BibitemShut
  {NoStop}%
\bibitem [{\citenamefont {Morse}\ \emph {et~al.}(2013)\citenamefont {Morse},
  \citenamefont {Orlov},\ and\ \citenamefont {Semertzidis}}]{Morse:2013hoa}%
  \BibitemOpen
  \bibfield  {author} {\bibinfo {author} {\bibfnamefont {W.~M.}\ \bibnamefont
  {Morse}}, \bibinfo {author} {\bibfnamefont {Y.~F.}\ \bibnamefont {Orlov}}, \
  and\ \bibinfo {author} {\bibfnamefont {Y.~K.}\ \bibnamefont {Semertzidis}},\
  }\href {\doibase 10.1103/PhysRevSTAB.16.114001} {\bibfield  {journal}
  {\bibinfo  {journal} {Phys.\ Rev.\ ST Accel.\ Beams}\ }\textbf {\bibinfo
  {volume} {16}},\ \bibinfo {pages} {114001} (\bibinfo {year}
  {2013})}\BibitemShut {NoStop}%
\bibitem [{\citenamefont {Rathmann}\ \emph {et~al.}(2020)\citenamefont
  {Rathmann}, \citenamefont {Nikolaev},\ and\ \citenamefont
  {Slim}}]{PhysRevAccelBeams.23.024601}%
  \BibitemOpen
  \bibfield  {author} {\bibinfo {author} {\bibfnamefont {F.}~\bibnamefont
  {Rathmann}}, \bibinfo {author} {\bibfnamefont {N.~N.}\ \bibnamefont
  {Nikolaev}}, \ and\ \bibinfo {author} {\bibfnamefont {J.}~\bibnamefont
  {Slim}},\ }\href {\doibase 10.1103/PhysRevAccelBeams.23.024601} {\bibfield
  {journal} {\bibinfo  {journal} {Phys. Rev. Accel. Beams}\ }\textbf {\bibinfo
  {volume} {23}},\ \bibinfo {pages} {024601} (\bibinfo {year}
  {2020})}\BibitemShut {NoStop}%
\bibitem [{\citenamefont {Maier}(1997)}]{Maier19971}%
  \BibitemOpen
  \bibfield  {author} {\bibinfo {author} {\bibfnamefont {R.}~\bibnamefont
  {Maier}},\ }\href {\doibase http://dx.doi.org/10.1016/S0168-9002(97)00324-0}
  {\bibfield  {journal} {\bibinfo  {journal} {Nuclear Instruments and Methods
  in Physics Research Section A: Accelerators, Spectrometers, Detectors and
  Associated Equipment}\ }\textbf {\bibinfo {volume} {390}},\ \bibinfo {pages}
  {1 } (\bibinfo {year} {1997})}\BibitemShut {NoStop}%
\bibitem [{\citenamefont {Weidemann}\ \emph {et~al.}(2015)\citenamefont
  {Weidemann} \emph {et~al.}}]{PAX}%
  \BibitemOpen
  \bibfield  {author} {\bibinfo {author} {\bibfnamefont {C.}~\bibnamefont
  {Weidemann}} \emph {et~al.},\ }\href {\doibase 10.1103/PhysRevSTAB.18.020101}
  {\bibfield  {journal} {\bibinfo  {journal} {Phys. Rev. ST Accel. Beams}\
  }\textbf {\bibinfo {volume} {18}},\ \bibinfo {pages} {020101} (\bibinfo
  {year} {2015})}\BibitemShut {NoStop}%
\bibitem [{\citenamefont {Wagner}\ \emph {et~al.}(2021)\citenamefont {Wagner},
  \citenamefont {Nass}, \citenamefont {Pretz}, \citenamefont {Abusaif},
  \citenamefont {Aggarwal}, \citenamefont {Andres}, \citenamefont {Bekman},
  \citenamefont {Canale}, \citenamefont {Ciepal}, \citenamefont {Ciullo},
  \citenamefont {Dahmen}, \citenamefont {Dymov}, \citenamefont {Ehrlich},
  \citenamefont {Gebel}, \citenamefont {Grigoryev}, \citenamefont {Grzonka},
  \citenamefont {Hejny}, \citenamefont {Hetzel}, \citenamefont {Kacharava},
  \citenamefont {Kamerdzhiev}, \citenamefont {Karanth}, \citenamefont
  {Keshelashvili}, \citenamefont {Kononov}, \citenamefont {Kulikov},
  \citenamefont {Laihem}, \citenamefont {Lehrach}, \citenamefont {Lenisa},
  \citenamefont {Lomidze}, \citenamefont {Magiera}, \citenamefont
  {Mchedlishvili}, \citenamefont {Müller}, \citenamefont {Nikolaev},
  \citenamefont {Pesce}, \citenamefont {Poncza}, \citenamefont {Rathmann},
  \citenamefont {Retzlaff}, \citenamefont {Saleev}, \citenamefont {Schmühl},
  \citenamefont {Shergelashvili}, \citenamefont {Shmakova}, \citenamefont
  {Slim}, \citenamefont {Stahl}, \citenamefont {Stephenson}, \citenamefont
  {Ströher}, \citenamefont {Tabidze}, \citenamefont {Tagliente}, \citenamefont
  {Talman}, \citenamefont {Uzikov}, \citenamefont {Valdau},\ and\ \citenamefont
  {Wrońska}}]{Wagner:2020akw}%
  \BibitemOpen
  \bibfield  {author} {\bibinfo {author} {\bibfnamefont {T.}~\bibnamefont
  {Wagner}}, \bibinfo {author} {\bibfnamefont {A.}~\bibnamefont {Nass}},
  \bibinfo {author} {\bibfnamefont {J.}~\bibnamefont {Pretz}}, \bibinfo
  {author} {\bibfnamefont {F.}~\bibnamefont {Abusaif}}, \bibinfo {author}
  {\bibfnamefont {A.}~\bibnamefont {Aggarwal}}, \bibinfo {author}
  {\bibfnamefont {A.}~\bibnamefont {Andres}}, \bibinfo {author} {\bibfnamefont
  {I.}~\bibnamefont {Bekman}}, \bibinfo {author} {\bibfnamefont
  {N.}~\bibnamefont {Canale}}, \bibinfo {author} {\bibfnamefont
  {I.}~\bibnamefont {Ciepal}}, \bibinfo {author} {\bibfnamefont
  {G.}~\bibnamefont {Ciullo}}, \bibinfo {author} {\bibfnamefont
  {F.}~\bibnamefont {Dahmen}}, \bibinfo {author} {\bibfnamefont
  {S.}~\bibnamefont {Dymov}}, \bibinfo {author} {\bibfnamefont
  {C.}~\bibnamefont {Ehrlich}}, \bibinfo {author} {\bibfnamefont
  {R.}~\bibnamefont {Gebel}}, \bibinfo {author} {\bibfnamefont
  {K.}~\bibnamefont {Grigoryev}}, \bibinfo {author} {\bibfnamefont
  {D.}~\bibnamefont {Grzonka}}, \bibinfo {author} {\bibfnamefont
  {V.}~\bibnamefont {Hejny}}, \bibinfo {author} {\bibfnamefont
  {J.}~\bibnamefont {Hetzel}}, \bibinfo {author} {\bibfnamefont
  {A.}~\bibnamefont {Kacharava}}, \bibinfo {author} {\bibfnamefont
  {V.}~\bibnamefont {Kamerdzhiev}}, \bibinfo {author} {\bibfnamefont
  {S.}~\bibnamefont {Karanth}}, \bibinfo {author} {\bibfnamefont
  {I.}~\bibnamefont {Keshelashvili}}, \bibinfo {author} {\bibfnamefont
  {A.}~\bibnamefont {Kononov}}, \bibinfo {author} {\bibfnamefont
  {A.}~\bibnamefont {Kulikov}}, \bibinfo {author} {\bibfnamefont
  {K.}~\bibnamefont {Laihem}}, \bibinfo {author} {\bibfnamefont
  {A.}~\bibnamefont {Lehrach}}, \bibinfo {author} {\bibfnamefont
  {P.}~\bibnamefont {Lenisa}}, \bibinfo {author} {\bibfnamefont
  {N.}~\bibnamefont {Lomidze}}, \bibinfo {author} {\bibfnamefont
  {A.}~\bibnamefont {Magiera}}, \bibinfo {author} {\bibfnamefont
  {D.}~\bibnamefont {Mchedlishvili}}, \bibinfo {author} {\bibfnamefont
  {F.}~\bibnamefont {Müller}}, \bibinfo {author} {\bibfnamefont
  {N.}~\bibnamefont {Nikolaev}}, \bibinfo {author} {\bibfnamefont
  {A.}~\bibnamefont {Pesce}}, \bibinfo {author} {\bibfnamefont
  {V.}~\bibnamefont {Poncza}}, \bibinfo {author} {\bibfnamefont
  {F.}~\bibnamefont {Rathmann}}, \bibinfo {author} {\bibfnamefont
  {M.}~\bibnamefont {Retzlaff}}, \bibinfo {author} {\bibfnamefont
  {A.}~\bibnamefont {Saleev}}, \bibinfo {author} {\bibfnamefont
  {M.}~\bibnamefont {Schmühl}}, \bibinfo {author} {\bibfnamefont
  {D.}~\bibnamefont {Shergelashvili}}, \bibinfo {author} {\bibfnamefont
  {V.}~\bibnamefont {Shmakova}}, \bibinfo {author} {\bibfnamefont
  {J.}~\bibnamefont {Slim}}, \bibinfo {author} {\bibfnamefont {A.}~\bibnamefont
  {Stahl}}, \bibinfo {author} {\bibfnamefont {E.}~\bibnamefont {Stephenson}},
  \bibinfo {author} {\bibfnamefont {H.}~\bibnamefont {Ströher}}, \bibinfo
  {author} {\bibfnamefont {M.}~\bibnamefont {Tabidze}}, \bibinfo {author}
  {\bibfnamefont {G.}~\bibnamefont {Tagliente}}, \bibinfo {author}
  {\bibfnamefont {R.}~\bibnamefont {Talman}}, \bibinfo {author} {\bibfnamefont
  {Y.}~\bibnamefont {Uzikov}}, \bibinfo {author} {\bibfnamefont
  {Y.}~\bibnamefont {Valdau}}, \ and\ \bibinfo {author} {\bibfnamefont
  {A.}~\bibnamefont {Wrońska}},\ }\href
  {https://dx.doi.org/10.1088/1748-0221/16/02/T02001} {\bibfield  {journal}
  {\bibinfo  {journal} {Journal of Instrumentation}\ }\textbf {\bibinfo
  {volume} {16}},\ \bibinfo {pages} {T02001} (\bibinfo {year}
  {2021})}\BibitemShut {NoStop}%
\bibitem [{\citenamefont {Wendt}(2021)}]{wendt2020bpm}%
  \BibitemOpen
  \bibfield  {author} {\bibinfo {author} {\bibfnamefont {M.}~\bibnamefont
  {Wendt}},\ }\href {\doibase 10.1109/MIM.2021.9620043} {\bibfield  {journal}
  {\bibinfo  {journal} {IEEE Instrumentation \& Measurement Magazine}\ }\textbf
  {\bibinfo {volume} {24}},\ \bibinfo {pages} {21} (\bibinfo {year}
  {2021})}\BibitemShut {NoStop}%
\bibitem [{\citenamefont {Slim}\ \emph {et~al.}(2016)\citenamefont {Slim},
  \citenamefont {Gebel}, \citenamefont {Heberling}, \citenamefont {Hinder},
  \citenamefont {H{\"o}lscher}, \citenamefont {Lehrach}, \citenamefont
  {Lorentz}, \citenamefont {Mey}, \citenamefont {Nass}, \citenamefont
  {Rathmann}, \citenamefont {Reifferscheidt}, \citenamefont {Soltner},
  \citenamefont {Straatmann}, \citenamefont {Trinkel},\ and\ \citenamefont
  {Wolters}}]{Slim2016116}%
  \BibitemOpen
  \bibfield  {author} {\bibinfo {author} {\bibfnamefont {J.}~\bibnamefont
  {Slim}}, \bibinfo {author} {\bibfnamefont {R.}~\bibnamefont {Gebel}},
  \bibinfo {author} {\bibfnamefont {D.}~\bibnamefont {Heberling}}, \bibinfo
  {author} {\bibfnamefont {F.}~\bibnamefont {Hinder}}, \bibinfo {author}
  {\bibfnamefont {D.}~\bibnamefont {H{\"o}lscher}}, \bibinfo {author}
  {\bibfnamefont {A.}~\bibnamefont {Lehrach}}, \bibinfo {author} {\bibfnamefont
  {B.}~\bibnamefont {Lorentz}}, \bibinfo {author} {\bibfnamefont
  {S.}~\bibnamefont {Mey}}, \bibinfo {author} {\bibfnamefont {A.}~\bibnamefont
  {Nass}}, \bibinfo {author} {\bibfnamefont {F.}~\bibnamefont {Rathmann}},
  \bibinfo {author} {\bibfnamefont {L.}~\bibnamefont {Reifferscheidt}},
  \bibinfo {author} {\bibfnamefont {H.}~\bibnamefont {Soltner}}, \bibinfo
  {author} {\bibfnamefont {H.}~\bibnamefont {Straatmann}}, \bibinfo {author}
  {\bibfnamefont {F.}~\bibnamefont {Trinkel}}, \ and\ \bibinfo {author}
  {\bibfnamefont {J.}~\bibnamefont {Wolters}},\ }\href {\doibase
  http://dx.doi.org/10.1016/j.nima.2016.05.012} {\bibfield  {journal} {\bibinfo
   {journal} {Nuclear Instruments and Methods in Physics Research Section A:
  Accelerators, Spectrometers, Detectors and Associated Equipment}\ }\textbf
  {\bibinfo {volume} {828}},\ \bibinfo {pages} {116 } (\bibinfo {year}
  {2016})}\BibitemShut {NoStop}%
\bibitem [{\citenamefont {Hinder}\ \emph {et~al.}(2016)\citenamefont {Hinder},
  \citenamefont {Krause}, \citenamefont {Soltner},\ and\ \citenamefont
  {Trinkel}}]{Hinder:2016hqi}%
  \BibitemOpen
  \bibfield  {author} {\bibinfo {author} {\bibfnamefont {F.}~\bibnamefont
  {Hinder}}, \bibinfo {author} {\bibfnamefont {H.-J.}\ \bibnamefont {Krause}},
  \bibinfo {author} {\bibfnamefont {H.}~\bibnamefont {Soltner}}, \ and\
  \bibinfo {author} {\bibfnamefont {F.}~\bibnamefont {Trinkel}},\ }in\ \href
  {https://accelconf.web.cern.ch/DOI/SRF2015/JACoW-SRF2015-TUPB015.html} {\emph
  {\bibinfo {booktitle} {{4th International Beam Instrumentation
  Conference}}}}\ (\bibinfo {year} {2016})\BibitemShut {NoStop}%
\bibitem [{\citenamefont {Abusaif}(2021)}]{phd_abusaif}%
  \BibitemOpen
  \bibfield  {author} {\bibinfo {author} {\bibfnamefont {F.}~\bibnamefont
  {Abusaif}},\ }\emph {\bibinfo {title} {Development of compact, highly
  sensitive beam position monitors for storage rings}},\ \href
  {https://www.doi.org/10.18154/RWTH-2021-10393} {Ph.D. thesis},\ \bibinfo
  {school} {RWTH Aachen University} (\bibinfo {year} {2021})\BibitemShut
  {NoStop}%
\bibitem [{\citenamefont {{D. Berners, L. Reginato}}(1992)}]{berners_rogowski}%
  \BibitemOpen
  \bibfield  {author} {\bibinfo {author} {\bibnamefont {{D. Berners, L.
  Reginato}}},\ }\href {https://doi.org/10.1063/1.44334} {\bibfield  {journal}
  {\bibinfo  {journal} {AIP Conf. Proc.}\ }\textbf {\bibinfo {volume} {281}},\
  \bibinfo {pages} {168 } (\bibinfo {year} {1992})}\BibitemShut {NoStop}%
\bibitem [{\citenamefont {Haci\"omeroglu}\ \emph {et~al.}(2019)\citenamefont
  {Haci\"omeroglu}, \citenamefont {Kawall}, \citenamefont {Lee}, \citenamefont
  {Matlashov}, \citenamefont {Omarov},\ and\ \citenamefont
  {Semertzidis}}]{Haciomeroglu:2018son}%
  \BibitemOpen
  \bibfield  {author} {\bibinfo {author} {\bibfnamefont {S.}~\bibnamefont
  {Haci\"omeroglu}}, \bibinfo {author} {\bibfnamefont {D.}~\bibnamefont
  {Kawall}}, \bibinfo {author} {\bibfnamefont {Y.-H.}\ \bibnamefont {Lee}},
  \bibinfo {author} {\bibfnamefont {A.}~\bibnamefont {Matlashov}}, \bibinfo
  {author} {\bibfnamefont {Z.}~\bibnamefont {Omarov}}, \ and\ \bibinfo {author}
  {\bibfnamefont {Y.~K.}\ \bibnamefont {Semertzidis}},\ }\href
  {https://www.doi.org/10.22323/1.340.0279} {\bibfield  {journal} {\bibinfo
  {journal} {PoS}\ }\textbf {\bibinfo {volume} {ICHEP2018}},\ \bibinfo {pages}
  {279} (\bibinfo {year} {2019})}\BibitemShut {NoStop}%
\bibitem [{\citenamefont {Rogowski}\ and\ \citenamefont
  {Steinhaus}(1912)}]{Rogowski1912}%
  \BibitemOpen
  \bibfield  {author} {\bibinfo {author} {\bibfnamefont {W.}~\bibnamefont
  {Rogowski}}\ and\ \bibinfo {author} {\bibfnamefont {W.}~\bibnamefont
  {Steinhaus}},\ }\href {\doibase 10.1007/BF01656479} {\bibfield  {journal}
  {\bibinfo  {journal} {Archiv f{\"u}r Elektrotechnik}\ }\textbf {\bibinfo
  {volume} {1}},\ \bibinfo {pages} {141} (\bibinfo {year} {1912})}\BibitemShut
  {NoStop}%
\bibitem [{\citenamefont {Nassisi}\ and\ \citenamefont
  {Luches}(1979)}]{doi:10.1063/1.1135946}%
  \BibitemOpen
  \bibfield  {author} {\bibinfo {author} {\bibfnamefont {V.}~\bibnamefont
  {Nassisi}}\ and\ \bibinfo {author} {\bibfnamefont {A.}~\bibnamefont
  {Luches}},\ }\href {\doibase 10.1063/1.1135946} {\bibfield  {journal}
  {\bibinfo  {journal} {Review of Scientific Instruments}\ }\textbf {\bibinfo
  {volume} {50}},\ \bibinfo {pages} {900} (\bibinfo {year} {1979})}\BibitemShut
  {NoStop}%
\bibitem [{\citenamefont {Pellinen}\ \emph {et~al.}(1980)\citenamefont
  {Pellinen}, \citenamefont {Di~Capua}, \citenamefont {Sampayan}, \citenamefont
  {Gerbracht},\ and\ \citenamefont {Wang}}]{doi:10.1063/1.1136119}%
  \BibitemOpen
  \bibfield  {author} {\bibinfo {author} {\bibfnamefont {D.~G.}\ \bibnamefont
  {Pellinen}}, \bibinfo {author} {\bibfnamefont {M.~S.}\ \bibnamefont
  {Di~Capua}}, \bibinfo {author} {\bibfnamefont {S.~E.}\ \bibnamefont
  {Sampayan}}, \bibinfo {author} {\bibfnamefont {H.}~\bibnamefont {Gerbracht}},
  \ and\ \bibinfo {author} {\bibfnamefont {M.}~\bibnamefont {Wang}},\ }\href
  {\doibase 10.1063/1.1136119} {\bibfield  {journal} {\bibinfo  {journal}
  {Review of Scientific Instruments}\ }\textbf {\bibinfo {volume} {51}},\
  \bibinfo {pages} {1535} (\bibinfo {year} {1980})},\ \Eprint
  {http://arxiv.org/abs/https://doi.org/10.1063/1.1136119}
  {https://doi.org/10.1063/1.1136119} \BibitemShut {NoStop}%
\bibitem [{\citenamefont {Han}\ \emph {et~al.}(2015)\citenamefont {Han},
  \citenamefont {Ding}, \citenamefont {Wu}, \citenamefont {Zhou}, \citenamefont
  {Jing}, \citenamefont {Liu}, \citenamefont {Chao},\ and\ \citenamefont
  {Qiu}}]{doi:10.1063/1.4916094}%
  \BibitemOpen
  \bibfield  {author} {\bibinfo {author} {\bibfnamefont {R.-Y.}\ \bibnamefont
  {Han}}, \bibinfo {author} {\bibfnamefont {W.-D.}\ \bibnamefont {Ding}},
  \bibinfo {author} {\bibfnamefont {J.-W.}\ \bibnamefont {Wu}}, \bibinfo
  {author} {\bibfnamefont {H.-B.}\ \bibnamefont {Zhou}}, \bibinfo {author}
  {\bibfnamefont {Y.}~\bibnamefont {Jing}}, \bibinfo {author} {\bibfnamefont
  {Q.-J.}\ \bibnamefont {Liu}}, \bibinfo {author} {\bibfnamefont {Y.-C.}\
  \bibnamefont {Chao}}, \ and\ \bibinfo {author} {\bibfnamefont {A.-C.}\
  \bibnamefont {Qiu}},\ }\href {\doibase 10.1063/1.4916094} {\bibfield
  {journal} {\bibinfo  {journal} {Review of Scientific Instruments}\ }\textbf
  {\bibinfo {volume} {86}},\ \bibinfo {pages} {035114} (\bibinfo {year}
  {2015})},\ \Eprint {http://arxiv.org/abs/https://doi.org/10.1063/1.4916094}
  {https://doi.org/10.1063/1.4916094} \BibitemShut {NoStop}%
\bibitem [{\citenamefont {Samimi}\ \emph {et~al.}(2015)\citenamefont {Samimi},
  \citenamefont {Mahari}, \citenamefont {Farahnakian},\ and\ \citenamefont
  {Mohseni}}]{samimi:2015}%
  \BibitemOpen
  \bibfield  {author} {\bibinfo {author} {\bibfnamefont {M.~H.}\ \bibnamefont
  {Samimi}}, \bibinfo {author} {\bibfnamefont {A.}~\bibnamefont {Mahari}},
  \bibinfo {author} {\bibfnamefont {M.~A.}\ \bibnamefont {Farahnakian}}, \ and\
  \bibinfo {author} {\bibfnamefont {H.}~\bibnamefont {Mohseni}},\ }\href
  {\doibase 10.1109/JSEN.2014.2362940} {\bibfield  {journal} {\bibinfo
  {journal} {IEEE Sensors Journal}\ }\textbf {\bibinfo {volume} {15}},\
  \bibinfo {pages} {651} (\bibinfo {year} {2015})}\BibitemShut {NoStop}%
\bibitem [{\citenamefont {Merzliakov}\ \emph {et~al.}(2016)\citenamefont
  {Merzliakov}, \citenamefont {B\"ohme},\ and\ \citenamefont
  {Kamerdzhiev}}]{ikpar2016:merzliakov}%
  \BibitemOpen
  \bibfield  {author} {\bibinfo {author} {\bibfnamefont {S.}~\bibnamefont
  {Merzliakov}}, \bibinfo {author} {\bibfnamefont {C.}~\bibnamefont {B\"ohme}},
  \ and\ \bibinfo {author} {\bibfnamefont {V.}~\bibnamefont {Kamerdzhiev}},\
  }\href
  {https://www.fz-juelich.de/de/ikp/downloads/jahresbericht_2016_artikel.pdf}
  {\bibfield  {journal} {\bibinfo  {journal} {IKP Annual Report,
  Forschungszentrum J\"ulich}\ } (\bibinfo {year} {2016})}\BibitemShut
  {NoStop}%
\bibitem [{\citenamefont {B\"ohme}\ \emph {et~al.}(2018)\citenamefont
  {B\"ohme}, \citenamefont {Bekman}, \citenamefont {Kamerdzhiev}, \citenamefont
  {Lorentz}, \citenamefont {Simon},\ and\ \citenamefont
  {Weidemann}}]{Bohme:2018sjy}%
  \BibitemOpen
  \bibfield  {author} {\bibinfo {author} {\bibfnamefont {C.}~\bibnamefont
  {B\"ohme}}, \bibinfo {author} {\bibfnamefont {I.}~\bibnamefont {Bekman}},
  \bibinfo {author} {\bibfnamefont {V.}~\bibnamefont {Kamerdzhiev}}, \bibinfo
  {author} {\bibfnamefont {B.}~\bibnamefont {Lorentz}}, \bibinfo {author}
  {\bibfnamefont {M.}~\bibnamefont {Simon}}, \ and\ \bibinfo {author}
  {\bibfnamefont {C.}~\bibnamefont {Weidemann}},\ }in\ \href
  {https://www.doi.org/10.18429/JACoW-IBIC2017-TUPCF20} {\emph {\bibinfo
  {booktitle} {{6th International Beam Instrumentation Conference}}}}\
  (\bibinfo {year} {2018})\BibitemShut {NoStop}%
\bibitem [{\citenamefont {Terman}(1943)}]{terman1943radio}%
  \BibitemOpen
  \bibfield  {author} {\bibinfo {author} {\bibfnamefont {F.}~\bibnamefont
  {Terman}},\ }\href {https://books.google.de/books?id=b7Q8AAAAIAAJ} {\emph
  {\bibinfo {title} {Radio Engineer's Handbook}}},\ McGraw-Hill handbooks\
  (\bibinfo  {publisher} {McGraw-Hill Book Company, Incorporated},\ \bibinfo
  {year} {1943})\BibitemShut {NoStop}%
\bibitem [{\citenamefont {{J.D. Jackson}}(1975)}]{jackson1975classical}%
  \BibitemOpen
  \bibfield  {author} {\bibinfo {author} {\bibnamefont {{J.D. Jackson}}},\
  }\href@noop {} {\emph {\bibinfo {title} {Classical Electrodynamics}}}\
  (\bibinfo  {publisher} {Wiley},\ \bibinfo {year} {1975})\BibitemShut
  {NoStop}%
\bibitem [{\citenamefont {Gradshteyn}\ \emph {et~al.}(2014)\citenamefont
  {Gradshteyn}, \citenamefont {Ryzhik}, \citenamefont {Zwillinger},\ and\
  \citenamefont {Moll}}]{Gradshteyn:1702455}%
  \BibitemOpen
  \bibfield  {author} {\bibinfo {author} {\bibfnamefont {I.~S.}\ \bibnamefont
  {Gradshteyn}}, \bibinfo {author} {\bibfnamefont {I.~M.}\ \bibnamefont
  {Ryzhik}}, \bibinfo {author} {\bibfnamefont {D.}~\bibnamefont {Zwillinger}},
  \ and\ \bibinfo {author} {\bibfnamefont {V.}~\bibnamefont {Moll}},\ }\href
  {https://www.sciencedirect.com/book/9780123849335/table-of-integrals-series-and-products}
  {\emph {\bibinfo {title} {{Table of integrals, series, and products; 8th
  ed.}}}}\ (\bibinfo  {publisher} {Academic Press},\ \bibinfo {address}
  {Amsterdam},\ \bibinfo {year} {2014})\BibitemShut {NoStop}%
\end{thebibliography}%

\end{document}